# Matrix-Variate Regressions and Envelope Models

Shanshan Ding

Department of Applied Economics and Statistics, University of Delaware

and

R. Dennis Cook

School of Statistics, University of Minnesota

**Abstract**

Modern technology often generates data with complex structures in which both response and explanatory variables are matrix-valued. Existing methods in the literature are able to tackle matrix-valued predictors but are rather limited for matrix-valued responses. In this article, we study matrix-variate regressions for such data, where the response $\mathbf{Y}$ on each experimental unit is a random matrix and the predictor $\mathbf{X}$ can be either a scalar, a vector, or a matrix, treated as non-stochastic in terms of the conditional distribution $\mathbf{Y}|\mathbf{X}$. We propose models for matrix-variate regressions and then develop envelope extensions of these models. Under the envelope framework, redundant variation can be eliminated in estimation and the number of parameters can be notably reduced when the matrix-variate dimension is large, possibly resulting in significant gains in efficiency. The proposed methods are applicable to high dimensional settings.

**Key Words:** Matrix-valued response; matrix-variate regression; reducing subspace; sufficient dimension reduction.

## 1 Introduction

Data with a matrix-valued response for each experimental unit are commonly encountered in contemporary statistical applications. For example, a longitudinal multivariate response can be treated integrally as a matrix-valued variable by designating rows and columns to be time and variates. Temporal and spatial data, multivariate growth curve data, image data and data from cross-over designs also generate matrix-valued responses. The methods in this article were motivated by data taken at different time points and different locations (or conditions), such as the twin cross-over bioassay study (Vϕlund, 1980) and EEG data (Li et al. 2010), where the measurements naturally form a matrix structure and can be treated as matrix-valued responses.

In these examples, the components of the matrix-variates are dependent among rows and columns. This dependence distinguishes longitudinal data in a matrix-valued response from traditional longitudinal modeling of vector-valued responses in which independent units are each measured over time. Vectorizing a matrix-valued response, or modeling the row or column vectors separately, typically loses dependency information and fails to capture the underlying data structure. Tackling matrix-variates directly can circumvent this issue. Research into this topic has gained considerable interest in recent years. Li et al. (2010) proposed a class of sufficient dimension reduction (SDR) methods, called dimension folding, for data with matrix-valued predictors. Pfeiffer et al. (2012) extended sliced inverse regression (SIR) to longitudinal predictors. Ding and Cook (2014, 2015a) developed model-based and model-free dimension folding methods for matrix-valued predictors. Xue and Yin (2014) introduced dimension folding SDR for conditional mean functions. More reviews can be found in Ding and Cook (2015b). On another track, Hung and Wang (2013), Zhou et al. (2013), and Zhou and Li (2014) extended generalized linear models (GLM) to matrix- and tensor-valued predictors for analyzing image data. All these methods, however, address data with matrix or tensor-valued predictors. Methods for dealing directly with matrix-valued responses are relatively limited in the literature. Viroli (2012) proposed matrix-variate regressions assuming independent rows for error matrices. The independence setting was assumed mainly for establishing theory but not for estimation. In applications modeling dependency among both rows and columns of a matrix-valued response can be desirable. Ding (2014) explored a general matrix-variate regression framework that allows such dependency and laid a foundation for the work in this article.

Li and Zhang (2015) studied tensor response regression where a linear relationship between a tensor-valued response and a predictor vector is considered. Their method for tensor responses is restricted to vector-valued predictors. Our motivation, model formulation, and estimation algorithms are different from Li and Zhang (2015). Firstly, our work was motivated by contemporary matrix-variate data and was geared to the development of matrix-variate analysis tools for such data. It is tailored for matrix-valued responses and can handle matrix-valued predictors simultaneously. Secondly, our method utilizes the matrix-variate structure and includes a novel parsimonious matrix linear model. Thus it can achieve parsimony even without an envelope structure, and can be applicable when the dimension of vec($\mathbf{X}$) is larger than the sample size $n$, while Li and Zhang' method does not achieve these goals. Our estimation procedure is also different from that used by Li and Zhang (2015). As discussed in Section 8 this can have important consequences in applications. Hoff (2015) studied multilinear tensor regression that can be used to model tensor responses and predictors including matrix settings. Yet theoretical properties were not fully investigated in the study. In addition, since tensor data often contain redundant information, without parsimonious modeling and estimation, the method might encounter less efficiency.

In this article, we first propose a novel matrix-variate regression in a general framework, where the response $\mathbf{Y}$ on each experimental unit is a random matrix and the predictor $\mathbf{X}$ can be either a scalar, a vector, or a matrix, treated as non-stochastic in terms of the conditional distribution $\mathbf{Y}|\mathbf{X}$. Neither the rows nor the columns of the matrix-valued variables are required to be independent to capture intrinsic dependent data structures. The method can also reduce the number of parameters and improve efficiency in estimation. Extraneous error variation may often occur beyond that anticipated by the model, particularly when the dimensions of the response are sizable. To allow for such situations, we further propose envelope methods for efficient estimation in matrix-variate regressions. One can then achieve dimension reduction in an analysis by eliminating redundant information, which can lead to substantial efficiency gains in estimation. While our matrix-variate regression and the correspondingly envelope model can be applied directly to high dimensional data, we also investigate sparse matrix-variate regression models to achieve enveloping with simultaneous matrix response variable selection.

The remainder of this article is organized as follows. In Section 2 we propose a new class of matrix-variate regression models and connect them with conventional regression models. Section 3 reviews the idea of enveloping and illustrates it with a real example. Section 4 is devoted to the development of envelope methods for matrix-variate regressions. Section 5 studies theoretical properties of matrix regression models and envelopes. Sparse versions of these models and corresponding estimation methods are discussed in Section 6. Sections 7 and 8 are devoted to illustrations with simulations and real data analyses. Technical details and additional results are included in a supplement. The codes for numerical studies, composed as an R software package named "matrixEnv", are available at https://sites.google.com/a/udel.edu/sding/software.

## 2 Matrix-variate regression

### 2.1 Model formulation

We consider modeling the regression of a matrix-valued response $\mathbf{Y} \in \mathbb{R}^{r \times m}$ on predictors that can be matrix-valued $\mathbf{X} \in \mathbb{R}^{p_1 \times p_2}$, vector-valued $\mathbf{X} \in \mathbb{R}^p$, or univariate, depending on the specific data. In the cross-over assay of insulin discussed in the Supplement Section P, the covariates are formed as a $2 \times 2$ matrix with elements indicating different treatment and dose levels. The goal is to investigate how the treatment and dose levels influence blood sugar concentration over time, a matrix-valued response.

As an obvious first step, we might use the vec operator that stacks the columns of a matrix to transform the problem into a standard vector-variate regression:

$$\text{vec}(\mathbf{Y}) = \boldsymbol{\mu} + \boldsymbol{\nu}\text{vec}(\mathbf{X}) + \boldsymbol{\epsilon}, \qquad (2.1)$$

where $\boldsymbol{\mu} \in \mathbb{R}^{rm}$, $\boldsymbol{\nu} \in \mathbb{R}^{rm \times p_1 p_2}$, and the error vector $\boldsymbol{\epsilon} \in \mathbb{R}^{rm}$ has mean 0 and covariance matrix $\boldsymbol{\Sigma} \in \mathbb{R}^{rm \times rm}$. In view of the potential for many parameters, this model may be hard to estimate well and difficult to interpret, unless perhaps some type of sparse structure is imposed. For problems in which a sparsity assumption is untenable, it seems to be a rather blunt instrument that neglects the matrix structure and may be grossly over-parameterized. We propose instead a relatively parsimoniously parameterized model that reflects the underlying structure of the matrix-variate response and predictor.

In preparation, the following operators for matrix-valued variables are predefined. For $\mathbf{Y} \in \mathbb{R}^{r \times m}$, the expectation of $\mathbf{Y}$ is $\text{E}(\mathbf{Y}) = \big(\text{E}(\mathbf{Y}_{ij})\big)$, the covariance matrix of $\mathbf{Y}$ is assumed to be of the form $\text{cov}[\text{vec}(\mathbf{Y})] = \boldsymbol{\Delta}_2 \otimes \boldsymbol{\Delta}_1$, where '$\otimes$' stands for the Kronecker product (De Waal 1985). Let $\text{cov}_r(\mathbf{Y}) = \text{E}[(\mathbf{Y} - \text{E}(\mathbf{Y}))^T(\mathbf{Y} - \text{E}(\mathbf{Y}))]$ be the covariance matrix over the rows of $\mathbf{Y}$ and let $\text{cov}_c(\mathbf{Y}) = \text{E}[(\mathbf{Y} - \text{E}(\mathbf{Y}))(\mathbf{Y} - \text{E}(\mathbf{Y}))^T]$ be the covariance matrix over the columns of $\mathbf{Y}$. Then as shown in Lemma 1 of the Supplement Section A, if $\text{cov}[\text{vec}(\mathbf{Y})]$ has the Kronecker structure, $\boldsymbol{\Delta}_2 = \text{cov}_r(Y)/\text{tr}(\boldsymbol{\Delta}_1)$ and $\boldsymbol{\Delta}_1 = \text{cov}_c(Y)/\text{tr}(\boldsymbol{\Delta}_2)$, where 'tr' means the trace operation. We thus similarly call $\boldsymbol{\Delta}_1$ and $\boldsymbol{\Delta}_2$ the column and row covariance matrices of $\mathbf{Y}$. They are uniquely defined up to a proportionality constant as $\boldsymbol{\Delta}_2 \otimes \boldsymbol{\Delta}_1 = a\boldsymbol{\Delta}_1 \otimes a^{-1}\boldsymbol{\Delta}_1$, where $a$ is a nonzero constant. We define the covariances between $\mathbf{Y}_1 \in \mathbb{R}^{r \times m}$ and $\mathbf{Y}_2 \in \mathbb{R}^{r \times m}$ in terms of the row covariance and column covariance, designated $\text{cov}_r(\mathbf{Y}_1, \mathbf{Y}_2) = \text{E}[(\mathbf{Y}_1 - \text{E}(\mathbf{Y}_1))^T(\mathbf{Y}_2 - \text{E}(\mathbf{Y}_2))]$ and $\text{cov}_c(\mathbf{Y}_1, \mathbf{Y}_2) = \text{E}[(\mathbf{Y}_1 - \text{E}(\mathbf{Y}_1))(\mathbf{Y}_2 - \text{E}(\mathbf{Y}_2))^T]$.

Given a matrix-variate predictor $\mathbf{X} \in \mathbb{R}^{p_1 \times p_2}$, we define the matrix regression of $\mathbf{Y} \in \mathbb{R}^{r \times m}$ on $\mathbf{X}$ as the bilinear model

$$\mathbf{Y} = \boldsymbol{\mu} + \boldsymbol{\beta}_1 \mathbf{X} \boldsymbol{\beta}_2^T + \boldsymbol{\varepsilon}, \qquad (2.2)$$

where $\boldsymbol{\mu} \in \mathbb{R}^{r \times m}$ is the overall mean, and $\boldsymbol{\beta}_1 \in \mathbb{R}^{r \times p_1}$ and $\boldsymbol{\beta}_2 \in \mathbb{R}^{m \times p_2}$ are the row and column coefficient matrices, which are uniquely defined only up to a proportionality constant. For identifiability, the column coefficient matrix $\boldsymbol{\beta}_2$ is defined to have Frobenius norm 1 and positive element in its first row and column. Different constraints mainly change the scale of the parameter matrices and have no effect in model fitting or prediction. The distribution of the matrix-valued random error $\boldsymbol{\varepsilon}$ is assumed be independent of $\mathbf{X}$, and have zero mean and covariance matrix $\text{cov}[\text{vec}(\boldsymbol{\varepsilon})] = \boldsymbol{\Sigma}_2 \otimes \boldsymbol{\Sigma}_1$, where as aforementioned, $\boldsymbol{\Sigma}_1 = \text{cov}_c(\boldsymbol{\varepsilon})/\text{tr}(\boldsymbol{\Sigma}_2)$ and $\boldsymbol{\Sigma}_2 = \text{cov}_r(\boldsymbol{\varepsilon})/\text{tr}(\boldsymbol{\Sigma}_1)$ are called the the column and row covariance matrices of $\boldsymbol{\varepsilon}$. Similarly, we require $\boldsymbol{\Sigma}_2$ to have unit Frobenius norm and positive first diagonal element in order to uniquely identify the two covariance matrices $\boldsymbol{\Sigma}_1$ and $\boldsymbol{\Sigma}_2$. If no constraints are required on $\boldsymbol{\beta}_2$ and $\boldsymbol{\Sigma}_2$, the Kronecker products $\boldsymbol{\beta}_2 \otimes \boldsymbol{\beta}_1$ and $\boldsymbol{\Sigma}_2 \otimes \boldsymbol{\Sigma}_1$ are still identifiable but the individual row and column parameter matrices are not.

The Kronecker covariance structure supposes a relational characteristic of the matrix-variate $\mathbf{Y}$, as the covariances of the column vectors of $\boldsymbol{\varepsilon}$ are all proportional to $\boldsymbol{\Sigma}_1$ and the covariances of the row vectors of $\boldsymbol{\varepsilon}$ are all proportional to $\boldsymbol{\Sigma}_2$. Such relationships are usually desirable for matrix-valued variables, especially for multivariate repeated measures and multivariate longitudinal data because of the intrinsic relationships among elements. For instance, the EEG data contains measurements of each subject from different time (row) and different scalp locations (column). It seems reasonable to model the data with similar variations among measurements over rows and measurements over columns. Another advantage of formulating the Kronecker covariance structure for such data is that the number of parameters in $\text{cov}[\text{vec}(\boldsymbol{\varepsilon})]$ can be dramatically reduced when the matrix dimension is high. When the Kronecker structure does not hold, a general covariance matrix $\text{cov}[\text{vec}(\boldsymbol{\varepsilon})] = \boldsymbol{\Sigma}$ can be applied. Hypothesis tests for the Kronecker covariance structure can be found in Shitan and Brockwell (1995), Lu and Zimmerman (2005), and Roy and Khattree (2005).

Let $\boldsymbol{e}_i$ denote the $m \times 1$ indicator vector with a 1 in the $i$-th element and zeros elsewhere, and imagine for the moment that $\boldsymbol{\beta}_2$ is known. Then the multivariate regression implied by (2.2) for the $j$-th column of $\mathbf{Y}$, $\mathbf{Y}\boldsymbol{e}_j = \boldsymbol{\mu}\boldsymbol{e}_j + \boldsymbol{\beta}_1 \mathbf{X} \boldsymbol{\beta}_2^T \boldsymbol{e}_j + \boldsymbol{\varepsilon}\boldsymbol{e}_j$, can be seen as standard multivariate regression with response vector $\mathbf{Y}\boldsymbol{e}_j$, coefficient matrix $\boldsymbol{\beta}_1$ and predictor vector $\mathbf{X}\boldsymbol{\beta}_2^T \boldsymbol{e}_j$. The matrix model (2.2) can be interpreted in the same way, except the predictors for each column of $\mathbf{Y}$ are estimated as linear combinations of the columns of $\mathbf{X}$ determined by the rows of $\boldsymbol{\beta}_2$. Hence the regression models for different columns of $\mathbf{Y}$ vary only through $\boldsymbol{\beta}_2$ and depend on different linear combinations of the columns of $\mathbf{X}$. The coefficients $\boldsymbol{\beta}_2$ can thus be seen to carry the column effect information. Similarly, the models for different rows of $\mathbf{Y}$ depend on different linear combinations of the rows of $\mathbf{X}$ and vary only via $\boldsymbol{\beta}_1$, which carries row effect information. As a consequence of this structure, model (2.2) can be applied in regressions where the sample size $n$ is smaller than the dimension of $\text{vec}(\mathbf{X})$. Further discussion is available in Section 2.3 and in the Supplement Section L.

Model (2.2) reduces the number of parameters by $\{rmp_1p_2 - (rp_1 + mp_2 - 1)\} + \{rm(rm + 1)/2 - r(r+1)/2 - m(m+1)/2 + 1\}$ in comparison to (2.1). The total number of the reduced parameters could be large when the matrix dimensions of $\mathbf{X}$ and $\mathbf{Y}$ are relatively high, leading to efficiency gains shown in Proposition 1, Section 5. Although reduced rank regression of $\text{vec}(\mathbf{Y})$ on $\text{vec}(\mathbf{X})$ can also reduce the number of parameters in estimation, like model (2.1) it does not employ any matrix structural information in model fitting. Thus, the method still belongs to the multivariate regression framework and in consequence $\boldsymbol{\nu}$ in (2.1) is not estimable when $n$ is smaller than the dimension of $\text{vec}(\mathbf{X})$ without imposing additional constraints. While outside the scope of this article, it is possible to extend the envelope structure proposed here to a reduced rank model for regression of $\text{vec}(\mathbf{Y})$ on $\text{vec}(\mathbf{X})$, following the development of reduced-rank envelope models by Cook et al. (2015).

In addition, the matrix regression (2.2) provides a general framework for regression modeling as it incorporates simple regression, multiple regression and multivariate multiple regression as special cases under different settings of $\mathbf{X}$ and $\mathbf{Y}$. For instance, when the response is univariate and the predictor is vector-valued, (2.2) reduces to a multiple regression. When both the response and the predictor are vector-valued, (2.2) coincides with the usual multivariate regression model. When $\mathbf{Y}$ is univariate and $\mathbf{X}$ is matrix-valued, $\text{E}(Y) = \mu + \boldsymbol{\beta}_1 \mathbf{X} \boldsymbol{\beta}_2^T = \mu + (\boldsymbol{\beta}_2 \otimes \boldsymbol{\beta}_1)\text{vec}(\mathbf{X})$. This rank-one formulation and its variants have been studied in recent work for modeling a univariate response with matrix-valued predictors (e.g., Ye et al., 2004; Hung and Wang, 2013; Zhou et al., 2013). Our approach encompasses the univariate response model as a special case and utilizes the matrix structural information of the predictors in modeling each element of the response with some sharing information for the same row/column elements.

Model (2.2) can be seen as a new regression tool for matrix-variate analyses in chemometrics. A variant on partial least square (PLS) regression is the primary method used to study matrix-variate problems in chemometrics (Smilde et al., 2004), which regresses both response and predictor on a common latent term with a bilinear coefficient structure. The relationship between the matrix-variate regression (2.2) and matrix PLS is similar to the relationship between regression and PLS in the conventional univariate and multivariate settings. Matrix structural relationships have also been shown to have practical value in other scenarios (e.g., Kong et al., 2005; Ye, 2005; Ding and Cook, 2014; Fausett and Fulton, 1994). As discussed in Sections 4 and 6, (2.2) can be further improved by employing the technique of enveloping for more efficiency gains, and by incorporating sparse structures for simultaneous variable selection.

Regressing $\mathbf{Y}$ on $\mathbf{X}$ or $\mathbf{X}^T$ can lead to different models in certain cases. Data with matrix-variate responses and predictors are often structured with rows and columns representing meaningful information and knowledge on matching $\mathbf{Y}$ with $\mathbf{X}$ or $\mathbf{X}^T$ may often be available. By using the commutation matrix $\mathbf{K}_{pq} \in \mathbb{R}^{pq \times pq}$, we have $\text{vec}(\mathbf{X}) = \mathbf{K}_{qp}\text{vec}(\mathbf{X}^T)$. For (2.1), as $\mathbf{K}_{qp}$ can be absorbed by the coefficient matrix $\boldsymbol{\nu}$, the model does not change when $\mathbf{X}$ is replaced by $\mathbf{X}^T$. For (2.2), we have $(\boldsymbol{\beta}_2 \otimes \boldsymbol{\beta}_1)\text{vec}(\mathbf{X}) = (\boldsymbol{\beta}_2 \otimes \boldsymbol{\beta}_1)\mathbf{K}_{qp}\text{vec}(\mathbf{X}^T)$. When either $\mathbf{Y}$ or $\mathbf{X}$ is vector-valued, $(\boldsymbol{\beta}_2 \otimes \boldsymbol{\beta}_1)K_{qp}$ is still a Kronecker product. Thus, (2.2) is not affected by using $\mathbf{X}$ or $\mathbf{X}^T$. However, $(\boldsymbol{\beta}_2 \otimes \boldsymbol{\beta}_1)\mathbf{K}_{qp}$ might not necessarily be a Kronecker product in general. In this case, choice between $\mathbf{X}$ and $\mathbf{X}^T$ may be clear in practice but if it is not, model comparison/selection with likelihood ratio test, AIC, BIC, or cross validation, can be useful to compare bilinear models using $\mathbf{X}$ and $\mathbf{X}^T$ individually with model (2.1).

Model (2.2) also encompasses several special formulations, such as concomitant columns (Viroli, 2012), univariate predictors, and the multivariate POD model (Cook and Weisberg, 2004). A detailed discussion is presented in the Supplement Section B.

## 2.2 Model estimation

In this section, we focus on the estimation of (2.2). Without a specific parametric distribution on the random error, one can estimate the coefficient parameters in (2.2) by using a loss function, like a squared loss function, and estimate the covariance matrices using moments. However, for statistical inference, as in conventional linear regressions, we assume that the random error $\boldsymbol{\varepsilon}$ follows a matrix normal distribution $N_{r \times m}(0, \boldsymbol{\Sigma}_1, \boldsymbol{\Sigma}_2)$. Background on the matrix normal distribution can be found in Dawid (1981) and De Waal (1985). We next describe maximum likelihood estimation of the parameters in (2.2) assuming the matrix normal. Details on the matrix normal density function and on the derivations of the maximum likelihood estimators (MLE) are given in the Supplement Section C.

Assume that $\mathbf{Y}_i$, $i = 1, \ldots, n$, are sampled from the conditional distributions $\mathbf{Y}|\mathbf{X}_i$, and that they are independent. Since $\mathbf{X}$ is non-stochastic we assume without loss of generality, that $\bar{\mathbf{X}} = 0$. For data analysis, we always center the predictors first before fitting the model. Then the MLE of $\boldsymbol{\mu}$ is $\bar{\mathbf{Y}}$. In matrix-variate analysis, explicit MLEs of $\boldsymbol{\beta}_1$, $\boldsymbol{\beta}_2$, $\boldsymbol{\Sigma}_1$ and $\boldsymbol{\Sigma}_2$ are generally unobtainable. We propose a two-step iterative algorithm to construct the estimators. We first hold $\boldsymbol{\beta}_2$ and $\boldsymbol{\Sigma}_2$ fixed and estimate the remaining parameters. We then use an iterative algorithm to obtain the full set of the MLEs. Let $\mathbf{C}_2 = \sum_{i=1}^{n}(\mathbf{Y}_i - \bar{\mathbf{Y}})\boldsymbol{\Sigma}_2^{-1}\boldsymbol{\beta}_2 \mathbf{X}_i^T$ and $\mathbf{M}_2 = \sum_{i=1}^{n} \mathbf{X}_i \boldsymbol{\beta}_2^T \boldsymbol{\Sigma}_2^{-1} \boldsymbol{\beta}_2 \mathbf{X}_i^T$. According to the log-likelihood function (C.1) in the Supplement, given $\boldsymbol{\beta}_2$ and $\boldsymbol{\Sigma}_2$ fixed, the estimators $\boldsymbol{\beta}_1$ and $\boldsymbol{\Sigma}_1$ are

$$\mathbf{B}_{1|2} = \mathbf{C}_2 \mathbf{M}_2^{-1} \quad \text{and} \quad \mathbf{S}_{\text{res}|2} = \frac{1}{nm}\sum_{i=1}^{n}(\mathbf{Y}_i - \bar{\mathbf{Y}} - \mathbf{B}_{1|2}\mathbf{X}_i\boldsymbol{\beta}_2^T)\boldsymbol{\Sigma}_2^{-1}(\mathbf{Y}_i - \bar{\mathbf{Y}} - \mathbf{B}_{1|2}\mathbf{X}_i\boldsymbol{\beta}_2^T)^T. \quad (2.3)$$

Similarly, let $\mathbf{C}_1 = \sum_{i=1}^{n}(\mathbf{Y}_i - \bar{\mathbf{Y}})^T \boldsymbol{\Sigma}_1^{-1} \boldsymbol{\beta}_1 \mathbf{X}_i$ and $\mathbf{M}_1 = \sum_{i=1}^{n} \mathbf{X}_i^T \boldsymbol{\beta}_1^T \boldsymbol{\Sigma}_1^{-1} \boldsymbol{\beta}_1 \mathbf{X}_i$. Given $\boldsymbol{\beta}_1$ and $\boldsymbol{\Sigma}_1$ fixed, $\boldsymbol{\beta}_2$ and $\boldsymbol{\Sigma}_2$ can be estimated by

$$\mathbf{B}_{2|1} = \mathbf{C}_1 \mathbf{M}_1^{-1} \quad \text{and} \quad \mathbf{S}_{\text{res}|1} = \frac{1}{nr}\sum_{i=1}^{n}(\mathbf{Y}_i^T - \bar{\mathbf{Y}}^T - \mathbf{B}_{2|1}\mathbf{X}_i^T\boldsymbol{\beta}_1^T)\boldsymbol{\Sigma}_1^{-1}(\mathbf{Y}_i^T - \bar{\mathbf{Y}}^T - \mathbf{B}_{2|1}\mathbf{X}_i^T\boldsymbol{\beta}_1^T)^T. \quad (2.4)$$

Let $\mathbf{B}_1$, $\mathbf{B}_2$, $\mathbf{S}_1$ and $\mathbf{S}_2$ be the MLEs of $\boldsymbol{\beta}_1$, $\boldsymbol{\beta}_2$, $\boldsymbol{\Sigma}_1$ and $\boldsymbol{\Sigma}_2$ respectively. Therefore, by initializing $\boldsymbol{\beta}_2 \in \mathbb{R}^{m \times q}$ with the required Frobenius normalization and initializing $\boldsymbol{\Sigma}_2 = n^{-1}\sum_{i=1}^{n}(\mathbf{Y}_i - \bar{\mathbf{Y}})^T(\mathbf{Y}_i - \bar{\mathbf{Y}})$, one can obtain $\mathbf{B}_1$, $\mathbf{B}_2$, $\mathbf{S}_1$ and $\mathbf{S}_2$ by iterating between (2.3) and (2.4) with updated values until the log-likelihood function of (2.2) meets a convergence criterion. Then normalize $\mathbf{B}_2$ as $\mathbf{B}_2 = \mathbf{B}_2/c$ and rescale $\mathbf{B}_1$ as $\mathbf{B}_1 = c\mathbf{B}_1$, where $c = \text{sign}(B_{11,2})\|\mathbf{B}_2\|_{\text{F}}$ with $B_{11,2}$ being the element in the first row and first column of $\mathbf{B}_2$ and '$\|\cdot\|_{\text{F}}$' representing the Frobenius norm; and normalize $\mathbf{S}_2$ as $\mathbf{S}_2 = \mathbf{S}_2/d$ and rescale $\mathbf{S}_1$ as $\mathbf{S}_1 = d\mathbf{S}_1$, where $d = \text{sign}(S_{11,2})\|\mathbf{S}_2\|_{\text{F}}$ and $S_{11,2}$ is the first diagonal element of $\mathbf{S}_2$.

If unique identification is not required for the individual parameter matrices $\boldsymbol{\beta}_j$s and $\boldsymbol{\Sigma}_j$s, the last step of normalization and rescaling in the algorithm is not necessary. In this case, we obtain MLEs for $\boldsymbol{\beta}_2 \otimes \boldsymbol{\beta}_1$ and $\boldsymbol{\Sigma}_2 \otimes \boldsymbol{\Sigma}_1$ but not the individual $\boldsymbol{\beta}_j$s and $\boldsymbol{\Sigma}_j$s. In addition, the aforementioned estimation procedure provides the same results as maximizing the likelihood directly under the normalization constraints on $\boldsymbol{\beta}_2$ and $\boldsymbol{\Sigma}_2$, because the corresponding MLEs with and without constraints are proportional. The justification is similar to the matrix normal estimation shown in Dutilleul (1999) and Srivastava et al. (2008).

The existence of the estimators $\mathbf{B}_1$, $\mathbf{B}_2$, $\mathbf{S}_1$ and $\mathbf{S}_2$ depends on the existence of the inverse matrices $\mathbf{S}_{\text{res}|1}^{-1}$ and $\mathbf{S}_{\text{res}|2}^{-1}$, which requires $n > \max(r/m, m/r)$ (Dutilleul, 1999), and on the existence of $\mathbf{M}_1^{-1}$ and $\mathbf{M}_2^{-1}$, which requires $p_1 \leq \min(p_2, m)n$ and $p_2 \leq \min(p_1, r)n$ respectively, when $\mathbf{B}_1$ and $\mathbf{B}_2$ are full rank matrices. These conditions seen fairly mild. They imply that the matrix-variate regression model can be directly applied to the high dimensional setting where $n < \dim(\text{vec}(\mathbf{X}))$, as the model estimation requires mainly

$$n \geq \max(r/m, m/r, p_1/\min(p_2, m), p_2/\min(p_1, r)).$$

A simulation study that assesses the performance of the matrix-variate regression in high dimensional settings is given in the Supplement Section L. In addition, to allow for possible sparse settings, we further propose sparse matrix-variate regressions in Section 6 to potentially improve estimation.

## 2.3 Goodness of fit

To guage how well the matrix regression (2.2) fits the observed data, one can test the goodness of fit of the model compared to an alternative. For instance, to compare (2.2) with the vector-variate regression of vec($\mathbf{Y}$) on vec($\mathbf{X}$) given in (2.1) assuming that $\boldsymbol{\epsilon} \sim N_{rm}(0, \boldsymbol{\Sigma})$, one can test the hypothesis: $H_0 : \boldsymbol{\nu} = \boldsymbol{\beta}_2 \otimes \boldsymbol{\beta}_1$, $\boldsymbol{\Sigma} = \boldsymbol{\Sigma}_2 \otimes \boldsymbol{\Sigma}_1$ versus $H_a$ : $H_0$ is not true. Since (2.2) is nested within (2.1), the likelihood ratio test (LRT) can be applied. Goodness of fit can also be evaluated by other model selection methods, such as AIC, BIC or cross validation. In the Supplement Section F, we numerically evaluated the goodness of fit of the matrix-variate regression (2.2) with LRT, AIC and BIC. All these methods were seen effective for identifying the true models. The goodness of fit of (2.2) compared to other models, such as to the special formulations in the Supplement Section B, can be similarly derived.

## 3 Review of envelopes

Since the envelope methodology introduced by Cook et al. (2010) is still relatively new, we provide a brief review in this section before turning to envelopes for matrix-variate regressions in Section 4. To facilitate the discussion, we introduce the following notations that will be used in the rest of the article. For an $r \times u$ matrix $\mathbf{B}$, span($\mathbf{B}$) is the subspace of $\mathbb{R}^r$ spanned by the columns of $\mathbf{B}$, $\mathbf{P}_\mathbf{B} = \mathbf{B}(\mathbf{B}^T\mathbf{B})^\dagger\mathbf{B}^T$ is the projection onto span($\mathbf{B}$), and $\mathbf{Q}_\mathbf{B} = \mathbf{I}_r - \mathbf{P}_\mathbf{B}$ is the orthogonal projection, where † is the Moore-Penrose inverse. A subspace $\mathcal{S} \subseteq \mathbb{R}^r$ is said to be a reducing subspace of $\mathbf{M} \in \mathbb{R}^{r \times r}$ if $\mathcal{S}$ decomposes $\mathbf{M}$ as $\mathbf{M} = \mathbf{P}_\mathcal{S} \mathbf{M} \mathbf{P}_\mathcal{S} + \mathbf{Q}_\mathcal{S} \mathbf{M} \mathbf{Q}_\mathcal{S}$, where $\mathbf{P}_\mathcal{S}$ is the projection onto $\mathcal{S}$ and $\mathbf{Q}_\mathcal{S}$ is the orthogonal projection. If $\mathcal{S}$ is a reducing subspace of $\mathbf{M}$, we say that $\mathcal{S}$ reduces $\mathbf{M}$.

Envelope methodology was proposed originally to improve efficiency in the vector-variate linear model

$$\mathbf{Y} = \boldsymbol{\alpha} + \boldsymbol{\beta} \mathbf{X} + \boldsymbol{\epsilon}, \tag{3.1}$$

where the response vector $\mathbf{Y} \in \mathbb{R}^r$, the predictor vector $\mathbf{X} \in \mathbb{R}^p$, $\boldsymbol{\alpha} \in \mathbb{R}^r$, $\boldsymbol{\beta} \in \mathbb{R}^{r \times p}$, and the random error $\boldsymbol{\epsilon} \sim N(0, \boldsymbol{\Sigma})$ is independent of $\mathbf{X}$. The motivation for enveloping in this context arose from asking if there are linear combinations of $\mathbf{Y}$ whose distribution is invariant to changes in the non-stochastic predictor $\mathbf{X}$. We refer to such linear combinations as $X$-invariants. To gain an operational version of this notion, suppose that the subspace $\mathcal{S} \subseteq \mathbb{R}^r$ satisfies the two conditions (i) the marginal distribution of $\mathbf{Q}_\mathcal{S} \mathbf{Y}$ does not depend on $\mathbf{X}$, and (ii) given the predictor $\mathbf{X}$, $\mathbf{P}_\mathcal{S} \mathbf{Y}$ and $\mathbf{Q}_\mathcal{S} \mathbf{Y}$ are independent. Then a change in $\mathbf{X}$ can affect the distribution of $\mathbf{Y}$ only via $\mathbf{P}_\mathcal{S} \mathbf{Y}$ (Cook et al. 2010). Informally, we think of $\mathbf{P}_\mathcal{S} \mathbf{Y}$ as the part of $\mathbf{Y}$ that is material to the regression, while $\mathbf{Q}_\mathcal{S} \mathbf{Y}$ is $X$-invariant and thus immaterial. Let $\mathcal{B} = \mathrm{span}(\boldsymbol{\beta})$. Cook et al. (2010) showed that the statistical conditions (i) and (ii) are equivalent to the algebraic conditions: (a) $\mathcal{B} \subseteq \mathcal{S}$, and (b) $\boldsymbol{\Sigma} = \mathbf{P}_\mathcal{S} \boldsymbol{\Sigma} \mathbf{P}_\mathcal{S} + \mathbf{Q}_\mathcal{S} \boldsymbol{\Sigma} \mathbf{Q}_\mathcal{S}$. Therefore, $\mathcal{S}$ is a reducing subspace of $\boldsymbol{\Sigma}$ that contains $\mathcal{B}$. The intersection of all reducing subspaces of $\boldsymbol{\Sigma}$ that contain $\mathcal{B}$ is called the $\boldsymbol{\Sigma}$-envelope of $\mathcal{B}$, denoted as $\mathcal{E}_{\boldsymbol{\Sigma}}(\mathcal{B})$, or $\mathcal{E}$ when used as a subscript. The envelope $\mathcal{E}_{\boldsymbol{\Sigma}}(\mathcal{B})$ serves to distinguish $\mathbf{P}_\mathcal{E} \mathbf{Y}$ and the maximal $X$-invariant $\mathbf{Q}_\mathcal{E} \mathbf{Y}$ in the estimation of $\boldsymbol{\beta}$ and can result in substantial increases in efficiency, sometimes equivalent to taking thousands of additional observations. Re-parameterizing (3.1) in terms of a semi-orthogonal basis matrix $\boldsymbol{\gamma}$ for $\mathcal{E}_{\boldsymbol{\Sigma}}(\mathcal{B})$, we have its envelope version $\mathbf{Y} = \boldsymbol{\alpha} + \boldsymbol{\gamma} \boldsymbol{\eta} \mathbf{X} + \boldsymbol{\epsilon}$ with $\boldsymbol{\Sigma} = \mathbf{P}_\mathcal{E} \boldsymbol{\Sigma} \mathbf{P}_\mathcal{E} + \mathbf{Q}_\mathcal{E} \boldsymbol{\Sigma} \mathbf{Q}_\mathcal{E}$. The standard errors of elements of the maximum likelihood estimator of $\boldsymbol{\beta} = \boldsymbol{\gamma} \boldsymbol{\eta}$ are often substantially smaller than the corresponding standard errors based on (3.1).

The overarching premise of the envelope $\mathcal{E}_{\boldsymbol{\Sigma}}(\mathcal{B})$ – that there may be linear combinations of $\mathbf{Y}$ whose distribution is invariant to changes in $\mathbf{X}$ – can be seen as inducing a type of generalized sparsity in both $\boldsymbol{\beta}$ and $\boldsymbol{\Sigma}$. Since $\mathcal{B} \in \mathcal{E}_{\boldsymbol{\Sigma}}(\mathcal{B})$, $\mathbf{Q}_\mathcal{E} \boldsymbol{\beta} = 0$ and thus linear combinations of $\boldsymbol{\beta}$ are 0 rather than individual elements. Envelopes go a step further by simultaneously requiring a generalized form of sparsity in $\boldsymbol{\Sigma} = \mathbf{P}_\mathcal{E} \boldsymbol{\Sigma} \mathbf{P}_\mathcal{E} + \mathbf{Q}_\mathcal{E} \boldsymbol{\Sigma} \mathbf{Q}_\mathcal{E}$.

We employ data on Concho water snakes (Johnson and Wichern, 2007) to illustrate the idea of enveloping intuitively. The data contain measurements on gender, age, tail length (mm), and snout to vent length (mm) for 37 female Concho water snakes and 29 male Concho water snakes. We want to investigate the gender effect on the tail length $Y_1$ and snout to vent length $Y_2$ for young snakes (age < 4 years). The data can be modeled using (3.1), where $\mathbf{Y} = (Y_1, Y_2)^T$ is a bivariate response vector and $X \in \mathbb{R}^1$ is an indicator variable with zero indicating a male snake and one indicating a female snake. Thus, the intercept $\boldsymbol{\alpha} \in \mathbb{R}^2$ is the mean of the male population, and the coefficient vector $\boldsymbol{\beta} \in \mathbb{R}^2$ is mean difference between the female population and the male population, which implies the gender effect. Figure 1(a) shows the data, where the circles represent the male snakes and the cross points represent the female snakes.

Under (3.1), inference on the gender difference in tail lengths proceeds by first projecting the data onto the horizontal axis of Figure 1(a) and then using univariate inference methods with the projected data. The two dashed curves represent the marginal distributions of the male tail length and the female tail length. These curves overlap substantially, so it may take a large sample size to detect the gender effect.

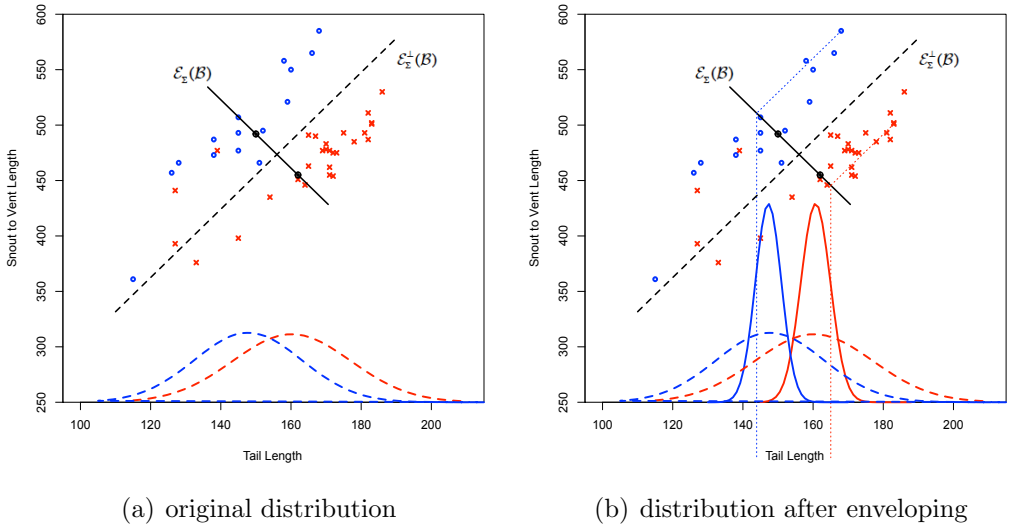

(a) original distribution     (b) distribution after enveloping

Figure 1: Illustration of enveloping.

The estimated envelope and its orthogonal complement are represented by the solid and the dashed lines in Figure 1(a). We can see that the two samples differ notably in the direction of the envelope, while differing little if at all in the direction of its orthogonal complement. Accordingly, envelope estimation proceeds by first projecting the data onto the envelope to remove $\mathbf{Q}_\mathcal{E}\mathbf{Y}$, the $X$-invariant part of $\mathbf{Y}$, and then projecting onto the horizontal axis to focus on the gender difference in tail length. This process is depicted in Figure 1(b), which shows the approximated marginal distributions of the male's and the female's tail lengths after enveloping, represented by the solid curves. The variances of two marginal distributions after enveloping are considerably reduced compared to the original ones (the dashed curves), which reflects the efficiency gain. The efficiency gains can be massive when the immaterial variation $\text{var}(\mathbf{Q}_\mathcal{E}\mathbf{Y})$ is substantially larger than the material variation $\text{var}(\mathbf{P}_\mathcal{E}\mathbf{Y}|X)$.

Recent developments of envelopes under other settings may be found in, for example, Su and Cook (2011, 2012), Cook et al. (2013), and Cook and Zhang (2015). In particular, Cook and Zhang (2015) studied the relationship between envelopes and canonical correlation analysis. They concluded that, while there is a weak population connection, canonical correlation analysis is not well-suited as a basis for envelope methodology.

# 4 Envelope models for matrix-variate regressions

Although the matrix-variate regression (2.2) is parsimoniously parameterized relative to the naive model (2.1), there still may be linear combinations of the rows or columns of $\mathbf{Y} \in \mathbb{R}^{p_1 \times p_2}$ whose distribution is invariant to changes in the predictors. To allow for the possibility of $X$-invariant linear combinations, we extend the vector-variate model described in Section 3 to matrix-variate regression. The rationale for doing so is generally the same as that described for vector-variate regressions: we hope to achieve efficiency in estimation and prediction better than that for model (2.2). We first establish the envelope structure for the matrix-variate regression model (2.2). Envelope models for the special cases are given in the Supplement.

## 4.1 Envelope formulation

To allow for the possibility that there are $X$-invariants in both the rows and columns of $\mathbf{Y}$, we suppose that there exist subspaces $\mathcal{S}_L \subseteq \mathbb{R}^r$ and $\mathcal{S}_R \subseteq \mathbb{R}^m$ so that

$$(a) \ \mathbf{Q}_{\mathcal{S}_L}\mathbf{Y} \mid \mathbf{X} \sim \mathbf{Q}_{\mathcal{S}_L}\mathbf{Y} \quad \text{and} \quad (b) \ \text{cov}_c(\mathbf{P}_{\mathcal{S}_L}\mathbf{Y}, \mathbf{Q}_{\mathcal{S}_L}\mathbf{Y} \mid \mathbf{X}) = 0 \qquad (4.1)$$

$$(a) \ \mathbf{Y}\mathbf{Q}_{\mathcal{S}_R} \mid \mathbf{X} \sim \mathbf{Y}\mathbf{Q}_{\mathcal{S}_R} \quad \text{and} \quad (b) \ \text{cov}_r(\mathbf{Y}\mathbf{P}_{\mathcal{S}_R}, \mathbf{Y}\mathbf{Q}_{\mathcal{S}_R} \mid \mathbf{X}) = 0, \qquad (4.2)$$

where $\text{cov}_c$ and $\text{cov}_r$ were defined in Section 2.1. These conditions are similar to conditions (i) and (ii) introduced in Section 3 for defining $X$-invariants of $\mathbf{Y}$ in vector-variate regressions. Let $\mathcal{B}_1 = \text{span}(\boldsymbol{\beta}_1)$ and $\mathcal{B}_2 = \text{span}(\boldsymbol{\beta}_2)$. Conditions (4.1a) and (4.2a) mean that the marginal distributions of $\mathbf{Q}_{\mathcal{S}_L}\mathbf{Y}$ and $\mathbf{Y}\mathbf{Q}_{\mathcal{S}_R}$ do not depend on $\mathbf{X}$, which is equivalent to requiring that $\mathcal{B}_1 \subseteq \mathcal{S}_L$ and $\mathcal{B}_2 \subseteq \mathcal{S}_R$. Conditions (4.1b) and (4.2b) indicate that $\mathbf{Q}_{\mathcal{S}_L}\mathbf{Y}$ does not respond to changes in $\mathbf{X}$ through a linear association with $\mathbf{P}_{\mathcal{S}_L}\mathbf{Y}$ and that $Y\mathbf{Q}_{\mathcal{S}_R}$ does not respond to changes in $\mathbf{X}$ through a linear association with $\mathbf{YP}_{\mathcal{S}_R}$. Further, condition (4.1b) holds if and only if $\mathbf{P}_{\mathcal{S}_L}\mathbf{\Sigma}_1\mathbf{Q}_{\mathcal{S}_L} = 0$, which is equivalent to requiring that $\mathcal{S}_L$ reduce $\mathbf{\Sigma}_1$, so $\mathbf{\Sigma}_1 = \mathbf{P}_{\mathcal{S}_L}\mathbf{\Sigma}_1\mathbf{P}_{\mathcal{S}_L}+\mathbf{Q}_{\mathcal{S}_L}\mathbf{\Sigma}_1\mathbf{Q}_{\mathcal{S}_L}$. Similarly, $\mathcal{S}_R$ must reduce $\mathbf{\Sigma}_2$, so $\mathbf{\Sigma}_2 = \mathbf{P}_{\mathcal{S}_R}\mathbf{\Sigma}_2\mathbf{P}_{\mathcal{S}_R}+\mathbf{Q}_{\mathcal{S}_R}\mathbf{\Sigma}_2\mathbf{Q}_{\mathcal{S}_R}$. Conditions (4.1b) and (4.2b) are weaker than the corresponding conditional independence conditions $\mathbf{P}_{\mathcal{S}_L}\mathbf{Y} \perp\!\!\!\perp \mathbf{Q}_{\mathcal{S}_L}\mathbf{Y} \mid \mathbf{X}$ and $\mathbf{YP}_{\mathcal{S}_R} \perp\!\!\!\perp \mathbf{YQ}_{\mathcal{S}_R} \mid \mathbf{X}$, paralleling to the idea of the envelope models of Cook et al. (2010). They are equivalent only under normality.

The intersection of all reducing subspaces $\mathcal{S}_L$ of $\mathbf{\Sigma}_1$ that contain $\mathcal{B}_1$ is the $\mathbf{\Sigma}_1$-envelope of $\mathcal{B}_1$, denoted as $\mathcal{E}_{\mathbf{\Sigma}_1}(\mathcal{B}_1)$ or $\mathcal{E}_1$ when used as a subscript. Similarly, the intersection of all reducing subspaces $\mathcal{S}_R$ of $\mathbf{\Sigma}_2$ that contain $\mathcal{B}_2$ is the $\mathbf{\Sigma}_2$-envelope of $\mathcal{B}_2$, denoted as $\mathcal{E}_{\mathbf{\Sigma}_2}(\mathcal{B}_2)$ or $\mathcal{E}_2$ when used as a subscript. These subspaces $\mathcal{E}_{\mathbf{\Sigma}_1}(\mathcal{B}_1)$ and $\mathcal{E}_{\mathbf{\Sigma}_2}(\mathcal{B}_2)$ always exist, are uniquely defined and serve as the fundamental constructs that allow row and column reduction of $\mathbf{Y}$. In concert, they imply that $\mathbf{Y}$ responds to changes in $\mathbf{X}$ only via $\mathbf{P}_{\mathcal{E}_1}\mathbf{YP}_{\mathcal{E}_2}$ and they hold the promise of much better estimation of $\boldsymbol{\beta}_1$ and $\boldsymbol{\beta}_2$. A theoretical justification regarding the variance reduction of enveloping is given by Proposition 4 in Section 5. Because the column spaces span($\boldsymbol{\beta}_1$) and span($\boldsymbol{\beta}_2$), and the orthogonal decompositions on $\mathbf{\Sigma}_1$ and $\mathbf{\Sigma}_2$, are all invariant under a multiplicative constant, the column and row envelopes $\mathcal{E}_1$ and $\mathcal{E}_2$ are always uniquely defined, whether normalization is required or not.

## 4.2 Envelope model

We next use $\mathcal{E}_{\mathbf{\Sigma}_1}(\mathcal{B}_1)$ and $\mathcal{E}_{\mathbf{\Sigma}_2}(\mathcal{B}_2)$ to re-parameterize (2.2) and to establish the envelope model. Let $\mathbf{L} \in \mathbb{R}^{r \times u_1}(u_1 \leq r)$ and $\mathbf{R} \in \mathbb{R}^{m \times u_2}(u_2 \leq m)$ be semi-orthogonal bases of $\mathcal{E}_{\mathbf{\Sigma}_1}(\mathcal{B}_1)$ and $\mathcal{E}_{\mathbf{\Sigma}_2}(\mathcal{B}_2)$, respectively, where $u_1$ and $u_2$ are the dimensions of the corresponding row and column envelopes and are temporarily assumed to be known. By definition, we know that span($\boldsymbol{\beta}_1$) $\subseteq \mathcal{E}_{\mathbf{\Sigma}_1}(\mathcal{B}_1)$ and span($\boldsymbol{\beta}_2$) $\subseteq \mathcal{E}_{\mathbf{\Sigma}_2}(\mathcal{B}_2)$, so there exist two coordinate matrices $\boldsymbol{\eta}_1 \in \mathbb{R}^{u_1 \times p_1}$ and $\boldsymbol{\eta}_2 \in \mathbb{R}^{u_2 \times p_2}$ such that $\boldsymbol{\beta}_1 = \mathbf{L}\boldsymbol{\eta}_1$ and $\boldsymbol{\beta}_2 = \mathbf{R}\boldsymbol{\eta}_2$. Let $(\mathbf{L}, \mathbf{L}_0)$ and $(\mathbf{R}, \mathbf{R}_0)$ be orthogonal matrices. Then the matrix regression model (2.2) can be re-parameterized as the following envelope model:

$$\mathbf{Y} = \boldsymbol{\mu} + \mathbf{L}\boldsymbol{\eta}_1\mathbf{X}\boldsymbol{\eta}_2^T\mathbf{R}^T + \boldsymbol{\varepsilon}, \tag{4.3}$$

where $\boldsymbol{\Sigma}_1 = \mathbf{L}\boldsymbol{\Omega}_1\mathbf{L}^T + \mathbf{L}_0\boldsymbol{\Omega}_{10}\mathbf{L}_0^T$, $\boldsymbol{\Sigma}_2 = \mathbf{R}\boldsymbol{\Omega}_2\mathbf{R}^T + \mathbf{R}_0\boldsymbol{\Omega}_{20}\mathbf{R}_0^T$, and $\boldsymbol{\Omega}_j > 0$ and $\boldsymbol{\Omega}_{j0} > 0$ are unknown, $j = 1, 2$. The re-parameterization of $\boldsymbol{\Sigma}_j$ is similar to the re-parameterization of $\boldsymbol{\beta}_j$. For example, because $\boldsymbol{\Sigma}_1$ has the decomposition form $\boldsymbol{\Sigma}_1 = \mathbf{P}_{\mathcal{S}_L}\boldsymbol{\Sigma}_1\mathbf{P}_{\mathcal{S}_L} + \mathbf{Q}_{\mathcal{S}_L}\boldsymbol{\Sigma}_1\mathbf{Q}_{\mathcal{S}_L}$, $\boldsymbol{\Omega}_1$ and $\boldsymbol{\Omega}_{10}$ then represent the coordinates of $\boldsymbol{\Sigma}_1$ relative to the basis $\mathbf{L}$ of $\mathcal{S}_L$ and its orthogonal basis $\mathbf{L}_0$, respectively. Similar description applies to $\boldsymbol{\Sigma}_2$.

As $\mathbf{L}\boldsymbol{\eta}_1$ and $\mathbf{R}\boldsymbol{\eta}_2$ are overparameterized, the matrices $\mathbf{L}$ and $\mathbf{R}$ themselves are not identifiable but their column spaces span($\mathbf{L}$) and span($\mathbf{R}$) are identifiable. The two parameter spaces are Grassmannians of dimension $u_1$ and $u_2$ in $\mathbb{R}^r$ and $\mathbb{R}^m$ with the numbers of unknown real parameters $u_1(r-u_1)$ and $u_2(m-u_2)$ respectively. Therefore, the total number of real parameters in (4.3) is $rm + u_1p_1 + u_2p_2 + r(r+1)/2 + m(m+1)/2 - 2$, which is equal to the sum of the numbers of parameters $rm$ in $\boldsymbol{\mu}$, $u_1p_1$ in $\boldsymbol{\eta}_1$, $u_2p_2 - 1$ in $\boldsymbol{\eta}_2$, $u_1(r - u_1)$ in $\mathbf{L}$, $u_2(m - u_2)$ in $\mathbf{R}$, $u_1(u_1+1)/2$ in $\boldsymbol{\Omega}_1$, $(r-u_1)(r-u_1+1)/2$ in $\boldsymbol{\Omega}_{10}$, and $u_2(u_2+1)/2 + (m-u_2)(m-u_2+1)/2 - 1$ in $\boldsymbol{\Omega}_2$ and $\boldsymbol{\Omega}_{20}$, whereas (2.2) has $rm + rp_1 + mp_2 + r(r+1)/2 + m(m+1)/2 - 2$ parameters. Here the normalization constraints on $\boldsymbol{\beta}_2$ and $\boldsymbol{\Sigma}_2$ eliminate two free parameters in each model. We see that the envelope model further reduces $(r - u_1)p_1 + (m - u_2)p_2$ parameters from (2.2).

Efficiency gain of envelope model (4.3) over matrix model (2.2) can arise in two distinct ways. The first is through parameter reduction, particularly when the number of real parameters in (2.2) is large relative to that in (4.3). But the largest efficiency gains are often realized when the variation in the $X$-invariant part of $\mathbf{Y}$ is large relative to the material variation $\mathbf{Y}$. Letting $\|\cdot\|_s$ denote the spectral norm of a matrix, this happens when $\|\boldsymbol{\Omega}_1\|_s \ll \|\boldsymbol{\Omega}_{10}\|_s$ or $\|\boldsymbol{\Omega}_2\|_s \ll \|\boldsymbol{\Omega}_{20}\|_s$. For vector-variate regression, this is reflected in Figure 1, since the variation along $\mathcal{E}_{\boldsymbol{\Sigma}}^\perp(\mathcal{B})$ is notably larger than that along $\mathcal{E}_{\boldsymbol{\Sigma}}(\mathcal{B})$.

## 4.3 Maximum likelihood estimation

In this section, we describe the MLEs for the unknown parameters in (4.3). As $\bar{\mathbf{X}} = 0$, the MLE of $\boldsymbol{\mu}$ is still $\bar{\mathbf{Y}}$ but the remaining parameter estimates cannot be expressed in closed form. Consequently, we propose a two-step iteration algorithm to obtain the MLEs.

Let $\hat{\mathbf{L}}$, $\hat{\mathbf{R}}$, $\hat{\boldsymbol{\beta}}_1$, $\hat{\boldsymbol{\beta}}_2$, $\hat{\boldsymbol{\Sigma}}_1$ and $\hat{\boldsymbol{\Sigma}}_2$ be the envelope MLEs of the corresponding parameters. We first hold $\boldsymbol{\beta}_2$ and $\boldsymbol{\Sigma}_2$ fixed and give the estimators of the remaining parameters. We later describe an iterative algorithm for computing the full set of MLEs. Assuming then that $\boldsymbol{\beta}_2$ and $\boldsymbol{\Sigma}_2$ are given, let $\mathbf{S}_{\mathbf{Y}|2} = (nm)^{-1}\sum_{i=1}^n (\mathbf{Y}_i - \bar{\mathbf{Y}})\boldsymbol{\Sigma}_2^{-1}(\mathbf{Y}_i - \bar{\mathbf{Y}})^T$, and let $\mathbf{B}_{1|2}$ and $\mathbf{S}_{\mathrm{res}|2}$ be defined in (2.3). Then $\mathbf{L}$ can be estimated as

$$\hat{\mathbf{L}} = \underset{\mathbf{G}}{\operatorname{argmin}} f_{\mathbf{L}|2}(\mathbf{G}), \tag{4.4}$$

where the objective function $f_{\mathbf{L}|2}(\mathbf{G}) = \log | \mathbf{G}^T \mathbf{S}_{\text{res}|2} \mathbf{G} | + \log | \mathbf{G}^T \mathbf{S}_{\mathbf{Y}|2}^{-1} \mathbf{G} |$ and $\operatorname{argmin}_G$ is taken over all semi-orthogonal matrices $\mathbf{G} \in \mathbb{R}^{r \times u_1}$. The estimators of $\boldsymbol{\beta}_1$ and $\boldsymbol{\Sigma}_1$ are then $\hat{\boldsymbol{\beta}}_{1|2} = \mathbf{P}_{\hat{\mathbf{L}}} \mathbf{B}_{1|2}$ and $\hat{\boldsymbol{\Sigma}}_{1|2} = \mathbf{P}_{\hat{\mathbf{L}}} \mathbf{S}_{\text{res}|2} \mathbf{P}_{\hat{\mathbf{L}}} + \mathbf{P}_{\hat{\mathbf{L}}_0} \mathbf{S}_{\mathbf{Y}|2} \mathbf{P}_{\hat{\mathbf{L}}_0}$, where $(\hat{\mathbf{L}}, \hat{\mathbf{L}}_0)$ is orthogonal. The detailed derivations are given in the Supplement Section C.

Similarly, given $\boldsymbol{\beta}_1$ and $\boldsymbol{\Sigma}_1$, let $\mathbf{S}_{\mathbf{Y}|1} = (nr)^{-1} \sum_{i=1}^n (\mathbf{Y}_i - \bar{\mathbf{Y}})^T \boldsymbol{\Sigma}_1^{-1} (\mathbf{Y}_i - \bar{\mathbf{Y}})$, and let $\mathbf{B}_{2|1}$ and $\mathbf{S}_{\text{res}|1}$ be defined in (2.4). Then

$$\hat{\mathbf{R}} = \underset{\mathbf{U}}{\operatorname{argmin}} f_{\mathbf{R}|1}(\mathbf{U}), \tag{4.5}$$

where $f_{\mathbf{R}|1} = \log | \mathbf{U}^T \mathbf{S}_{\text{res}|1} \mathbf{U} | + \log | \mathbf{U}^T \mathbf{S}_{\mathbf{Y}|1}^{-1} \mathbf{U} |$ and $\operatorname{argmin}_{\mathbf{U}}$ is taken over all semi-orthogonal matrices $\mathbf{U} \in \mathbb{R}^{m \times u_2}$. Then the intermediate estimators of $\boldsymbol{\beta}_2$ and $\boldsymbol{\Sigma}_2$ are $\hat{\boldsymbol{\beta}}_{2|1} = \mathbf{P}_{\hat{\mathbf{R}}} \mathbf{B}_{2|1}$ and $\hat{\boldsymbol{\Sigma}}_{2|1} = \mathbf{P}_{\hat{\mathbf{R}}} \mathbf{S}_{\text{res}|1} \mathbf{P}_{\hat{\mathbf{R}}} + \mathbf{P}_{\hat{\mathbf{R}}_0} \mathbf{S}_{\mathbf{Y}|1} \mathbf{P}_{\hat{\mathbf{R}}_0}$, where $(\hat{\mathbf{R}}, \hat{\mathbf{R}}_0)$ is orthogognal. The full MLEs can now be obtained by alternating between $(\hat{\boldsymbol{\beta}}_{2|1}, \hat{\boldsymbol{\Sigma}}_{2|1})$ and $(\hat{\boldsymbol{\beta}}_{1|2}, \hat{\boldsymbol{\Sigma}}_{1|2})$, and then rescaling them:

1. Initialize $\boldsymbol{\beta}_{20}$ and $\boldsymbol{\Sigma}_{20}$ as the MLEs from (2.2). Let $\hat{\boldsymbol{\beta}}_2 = \boldsymbol{\beta}_{20}$ and let $\hat{\boldsymbol{\Sigma}}_2 = \boldsymbol{\Sigma}_{20}$.

2. Given $\hat{\boldsymbol{\beta}}_2$ and $\hat{\boldsymbol{\Sigma}}_2$, estimate $\mathbf{L}$ by (4.4), $\hat{\boldsymbol{\beta}}_1 = \mathbf{P}_{\hat{\mathbf{L}}} \mathbf{B}_{1|2}$ and $\hat{\boldsymbol{\Sigma}}_1 = \mathbf{P}_{\hat{\mathbf{L}}} \mathbf{S}_{\text{res}|2} \mathbf{P}_{\hat{\mathbf{L}}} + \mathbf{P}_{\hat{\mathbf{L}}_0} \mathbf{S}_{\mathbf{Y}|2} \mathbf{P}_{\hat{\mathbf{L}}_0}$.

3. Given $\hat{\boldsymbol{\beta}}_1$ and $\hat{\boldsymbol{\Sigma}}_1$, estimate $\mathbf{R}$ by (4.5), $\hat{\boldsymbol{\beta}}_2 = \mathbf{P}_{\hat{\mathbf{R}}} \mathbf{B}_{2|1}$ and $\hat{\boldsymbol{\Sigma}}_2 = \mathbf{P}_{\hat{\mathbf{R}}} \mathbf{S}_{\text{res}|1} \mathbf{P}_{\hat{\mathbf{R}}} + \mathbf{P}_{\hat{\mathbf{R}}_0} \mathbf{S}_{\mathbf{Y}|1} \mathbf{P}_{\hat{\mathbf{R}}_0}$.

4. Iterate 2-3 until the log-likelihood function of (4.3) converges. Then normalize $\hat{\boldsymbol{\beta}}_2$ as $\hat{\boldsymbol{\beta}}_2 = \hat{\boldsymbol{\beta}}_2/c$ and rescale $\hat{\boldsymbol{\beta}}_1$ as $\hat{\boldsymbol{\beta}}_1 = c\hat{\boldsymbol{\beta}}_1$, where $c = \operatorname{sign}(\hat{\beta}_{11,2}) \|\hat{\boldsymbol{\beta}}_2\|_F$ and $\hat{\beta}_{11,2}$ is the first element of $\operatorname{vec}(\hat{\boldsymbol{\beta}}_2)$, and normalize $\hat{\boldsymbol{\Sigma}}_2$ as $\hat{\boldsymbol{\Sigma}}_2 = \hat{\boldsymbol{\Sigma}}_2/d$ and rescale $\hat{\boldsymbol{\Sigma}}_1$ as $\hat{\boldsymbol{\Sigma}}_1 = d\hat{\boldsymbol{\Sigma}}_1$, where $d = \operatorname{sign}(\hat{\Sigma}_{11,2}) \|\hat{\boldsymbol{\Sigma}}_2\|_F$ and $\hat{\Sigma}_{11,2}$ is the first diagonal element of $\hat{\boldsymbol{\Sigma}}_2$.

Similar to the discussion in Section 2.2, the proposed estimation method provides the same results as estimating the parameters directly under constrained $\boldsymbol{\beta}_2$ and $\boldsymbol{\Sigma}_2$. If one is interested only in estimating the Kronecker products $\boldsymbol{\beta}_2 \otimes \boldsymbol{\beta}_1$ and $\boldsymbol{\Sigma}_2 \otimes \boldsymbol{\Sigma}_1$, but not individual $\boldsymbol{\beta}_j$s and $\boldsymbol{\Sigma}_j$s, the last rescaling step in the algorithm is unnecessary. We optimize $f_{\mathbf{L}|2}(\mathbf{G})$ and $f_{\mathbf{R}|1}(\mathbf{U})$ by using the non-Grassman algorithm proposed by Cook et al. (2016) that is more efficient and faster than existing Grassmannian optimization approaches. In particular, Cook et al. (2016) demonstrated its superiority over the 1D algorithm applied by Li and Zhang (2015).

The resulting envelope estimators $\hat{\boldsymbol{\beta}}_1 = \mathbf{P}_{\hat{\mathbf{L}}}\mathbf{B}_{1|2}$ and $\hat{\boldsymbol{\beta}}_2 = \mathbf{P}_{\hat{\mathbf{R}}}\mathbf{B}_{2|1}$ in Steps 2 and 3 can be viewed as the projections of their corresponding estimators, $\mathbf{B}_{1|2}$ and $\mathbf{B}_{2|1}$, from (2.2) onto the estimated row and column envelopes $\hat{\mathcal{E}}_{\boldsymbol{\Sigma}_1}(\mathcal{B}_1)$ and $\hat{\mathcal{E}}_{\boldsymbol{\Sigma}_2}(\mathcal{B}_2)$. In addition, the envelope MLEs $\hat{\boldsymbol{\Sigma}}_1$ and $\hat{\boldsymbol{\Sigma}}_2$ are both partitioned into the estimated $\mathbf{X}$-variant and $X$-invariant parts of $\mathbf{Y}$. Similar to the intuitive arguments in Figure 1, with the envelope projections, the formulations here can lead to more efficient estimation of the parameters. To accommodate possible sparse structures, sparse envelope matrix regressions will be further investigated in Section 6 for simultaneous variable selection.

So far, we have assumed that the dimensions $u_1$ and $u_2$ of the row and column envelopes are known. Extending the methods described by Cook et al. (2010), these dimensions, which are essentially model selection parameters, can be determined by using an information criterion, say AIC or BIC, or using other model selection methods such as cross validation and likelihood ratio tests. For example, for AIC and BIC, the envelope dimensions $u_1$ and $u_2$ can be selected by minimizing the objective function $-2\hat{l}(u_1, u_2) + h(n)t(u_1, u_2)$, where $\hat{l}(u_1, u_2)$ is the maximized log-likelihood function of the envelope matrix regression model at the indicated dimensions, which is given at (C.2) in the Supplement, $h(n)$ is equal to $\log(n)$ for BIC and is equal to 2 for AIC, and $t(u_1, u_2) = rm + u_1 p_1 + u_2 p_2 + r(r+1)/2 + m(m+1)/2 - 2$ is the total number of parameters in the target model. We evaluated the numerical performance of the proposed dimension selection methods in the Supplement Section N, which show fairly accurate results. For data analysis, we mainly focused on BIC for envelope dimension selection. Examples are given in numerical studies in Section 8.

## 5 Theoretical properties

In this section, we investigate the asymptotic properties of the MLEs from (2.2) and (4.3). We show that the MLEs from (2.2) are asymptotically more efficient than the MLEs obtained from (2.1) and that the envelope estimators from (4.3) can gain efficiency over the MLEs from (2.2).

We neglect $\boldsymbol{\mu}$ in the theoretical development, since its MLE is $\bar{\mathbf{Y}}$ in all the three models, which is asymptotically independent of the MLEs of all other parameters in the models. To show the relative efficiency of (2.2) to (2.1), we need to consider only the joint asymptotic distribution of the MLEs of $\text{vec}(\boldsymbol{\beta}_2 \otimes \boldsymbol{\beta}_1)$ and $\text{vech}(\boldsymbol{\Sigma}_2 \otimes \boldsymbol{\Sigma}_1)$, whereas the individual parameter matrices $\boldsymbol{\beta}_j$s and $\boldsymbol{\Sigma}_j$s are not of interest. In this case, the matrix regression (2.2) can be considered as an over-parameterized structural model of (2.1). Hence we can apply Proposition 4.1 in Shapiro (1986) to derive the target property. For notation convenience, let $\boldsymbol{\gamma} = (\text{vec}(\boldsymbol{\nu})^T, \text{vech}(\boldsymbol{\Sigma})^T)^T$ and $\boldsymbol{\theta} = (\text{vec}(\boldsymbol{\beta}_1)^T, \text{vec}(\boldsymbol{\beta}_2)^T, \text{vech}(\boldsymbol{\Sigma}_1)^T, \text{vech}(\boldsymbol{\Sigma}_2)^T)^T$, where 'vech' denotes the half-vectorization operator. Then under (2.2), we can rewrite $\boldsymbol{\gamma}$ as $\boldsymbol{\gamma} = (h_1(\boldsymbol{\theta})^T, h_2(\boldsymbol{\theta})^T)^T := h(\boldsymbol{\theta})$, where $h_1(\boldsymbol{\theta}) = \text{vec}(\boldsymbol{\beta}_2 \otimes \boldsymbol{\beta}_1)$ and $h_2(\boldsymbol{\theta}) = \text{vech}(\boldsymbol{\Sigma}_2 \otimes \boldsymbol{\Sigma}_1)$. Define $\hat{\boldsymbol{\gamma}}_s$ to be the standard MLE of $\boldsymbol{\gamma}$ from (2.1) and define $\hat{\boldsymbol{\gamma}}_m$ to be the MLE of $\boldsymbol{\gamma}$ from the matrix regression (2.2). For a general statistic $\mathbf{T}_n$, denote the asymptotic variance of $\sqrt{n}(\mathbf{T}_n - \boldsymbol{\theta})$ as $\text{avar}(\sqrt{n}\mathbf{T}_n)$. Let $\mathbf{J}$ be the Fisher information function of $\boldsymbol{\gamma}$ and let $\mathbf{H} = (\partial h_i/\partial \boldsymbol{\theta}_j)_{i,j}$ be the gradient matrix. Proposition 1 establishes the asymptotic distribution of $\hat{\boldsymbol{\gamma}}_m$ and its asymptotic efficiency relative to $\hat{\boldsymbol{\gamma}}_s$. The proof is given in the Supplement Section G.

**Proposition 1.** *Under (2.2) with a normal error, $\sqrt{n}(\hat{\boldsymbol{\gamma}}_m - \boldsymbol{\gamma})$ converges in distribution to a normal random vector with mean zero and covariance matrix $\text{avar}(\sqrt{n}\hat{\boldsymbol{\gamma}}_m) = \mathbf{H}(\mathbf{H}^T\mathbf{J}\mathbf{H})^\dagger\mathbf{H}^T$. Moreover, $\text{avar}(\sqrt{n}\hat{\boldsymbol{\gamma}}_m) \leq \text{avar}(\sqrt{n}\hat{\boldsymbol{\gamma}}_s)$.*

Consequently, when the matrix regression structure holds, omitting row and column information in the coefficient and covariance formulation can result in loss of efficiency in parameter estimation. We next establish asymptotic properties for the MLE from (4.3) and show that enveloping can further improve efficiency in estimating (2.2). Let

$$\boldsymbol{\zeta} = (\text{vec}(\boldsymbol{\eta}_1)^T, \text{vec}(\mathbf{L})^T, \text{vec}(\boldsymbol{\eta}_2)^T, \text{vec}(\mathbf{R})^T, \text{vech}(\boldsymbol{\Omega}_1)^T, \text{vech}(\boldsymbol{\Omega}_{10})^T, \text{vech}(\boldsymbol{\Omega}_2)^T, \text{vech}(\boldsymbol{\Omega}_{20})^T)^T$$

be the collection of the envelope parameters. Under (4.3), $\boldsymbol{\theta}$ is reformulated as

$$\boldsymbol{\theta} = (g_1(\boldsymbol{\zeta})^T, g_2(\boldsymbol{\zeta})^T, g_3(\boldsymbol{\zeta})^T, g_4(\boldsymbol{\zeta})^T)^T := g(\boldsymbol{\zeta}),$$

where $g_1(\boldsymbol{\zeta}) = \text{vec}(\mathbf{L}\boldsymbol{\eta}_1)$, $g_2(\boldsymbol{\zeta}) = \text{vec}(\mathbf{R}\boldsymbol{\eta}_2)$, $g_3(\boldsymbol{\zeta}) = \text{vech}(\mathbf{L}\boldsymbol{\Omega}_1\mathbf{L}^T + \mathbf{L}_0\boldsymbol{\Omega}_{10}\mathbf{L}_0^T)$, and $g_4(\boldsymbol{\zeta}) = \text{vech}(\mathbf{R}\boldsymbol{\Omega}_2\mathbf{R}^T + \mathbf{R}_0\boldsymbol{\Omega}_{20}\mathbf{R}_0^T)$. Thus, $\boldsymbol{\gamma} = h(g(\boldsymbol{\zeta})) := \phi(\boldsymbol{\zeta})$. Correspondingly, the envelope MLE of $\boldsymbol{\gamma}$ is $\phi(\hat{\boldsymbol{\zeta}})$. Let $\mathbf{G}_1 = (\partial g_i/\partial \boldsymbol{\zeta}_j)_{i,j}$ be a gradient matrix.

**Proposition 2.** *Under (4.3) with a normal error, $\sqrt{n}(\phi(\hat{\boldsymbol{\zeta}}) - \phi(\boldsymbol{\zeta}))$ converges in distribution to a normal random vector with mean zero and covariance matrix $\text{avar}(\sqrt{n}\phi(\hat{\boldsymbol{\zeta}})) = \mathbf{H}\mathbf{G}_1(\mathbf{G}_1^T\mathbf{H}^T\mathbf{J}\mathbf{H}\mathbf{G}_1)^\dagger \mathbf{G}_1^T\mathbf{H}^T$. Moreover, $\text{avar}(\sqrt{n}\phi(\hat{\boldsymbol{\zeta}})) \leq \text{avar}(\sqrt{n}\hat{\boldsymbol{\gamma}}_m)$.*

The proof is given in the Supplement Section G. So far we have derived the asymptotic distributions for the estimators of $\boldsymbol{\gamma}$ under different model settings and compared the relative efficiency among different estimators. We next investigate the asymptotic properties of the individual row and column parameter estimators of $\boldsymbol{\theta}$ in (2.2) and (4.3), given that these parameters are uniquely defined by normalization as discussed in the model settings. Let $\hat{\boldsymbol{\theta}}$ denote the MLE of $\boldsymbol{\theta}$ from (2.2), and let $g(\hat{\boldsymbol{\zeta}})$ denote the envelope MLE of $\boldsymbol{\theta}$ from (4.3).

**Proposition 3.** *Under (2.2) with a normal error and uniquely defined row and column parameter matrices, $\sqrt{n}(\hat{\boldsymbol{\theta}} - \boldsymbol{\theta})$ converges in distribution to a normal random vector with mean zero and covariance matrix $\mathrm{avar}(\sqrt{n}\hat{\boldsymbol{\theta}}) = \mathbf{K}\mathbf{J}_1^{-1}\mathbf{K}^T$, where $\mathbf{K}$ and $\mathbf{J}_1$ are matrix-valued functions of $\boldsymbol{\theta}$ given in the Supplement Section H.*

**Proposition 4.** *Under (4.3) with a normal error and uniquely defined row and column parameter matrices, $\sqrt{n}(g(\hat{\boldsymbol{\zeta}}) - g(\boldsymbol{\zeta}))$ converges in distribution to a normal random vector with mean zero and covariance matrix $\mathrm{avar}(\sqrt{n}g(\hat{\boldsymbol{\zeta}})) = \mathbf{K}\boldsymbol{G}_1(\boldsymbol{G}_1^T\mathbf{J}_1\boldsymbol{G}_1)^\dagger\boldsymbol{G}_1^T\mathbf{K}$. Moreover, $\mathrm{avar}(\sqrt{n}g(\hat{\boldsymbol{\zeta}})) \leq \mathrm{avar}(\sqrt{n}\hat{\boldsymbol{\theta}})$.*

Propositions 3 and 4 demonstrate the asymptotic normality for both enveloped and non-enveloped estimators of $\boldsymbol{\theta}$. Proposition 4 further indicates that as long as there exists $X$-invariants in (2.2), one can gain potential efficiency in estimation by enveloping the matrix regression model. The proofs are given in the Supplement Section H. Since the row and column coefficient matrices, $\mathrm{vec}(\boldsymbol{\beta}_1)$ and $\mathrm{vec}(\boldsymbol{\beta}_2)$, are often of interest in regression analysis, we further derived asymptotic variances for $\mathrm{vec}(\hat{\boldsymbol{\beta}}_1)$ and $\mathrm{vec}(\hat{\boldsymbol{\beta}}_2)$ in the Supplement Section I.

When the random error in (2.2) is not normal, the estimator of $\boldsymbol{\theta}$ under a type of generalized least square estimation (as shown in the Supplement Section J) is the same as the MLE of (2.2) under normality. We thus use the same notation $\hat{\boldsymbol{\theta}}$. Under mild conditions, the estimator $\hat{\boldsymbol{\theta}}$ is still asymptotically normal, except having a more complex asymptotic covariance matrix. The terms $\mathbf{K}$ and $\mathbf{J}_1$ described in Proposition 3 will be used in the following propositions. The proofs of the propositions are given in the Supplement Section J.

**Proposition 5.** *Under (2.2) with uniquely defined row and column parameter matrices, suppose that the random error $\boldsymbol{\varepsilon}$ follows a general matrix-valued distribution with zero mean and column and row covariance matrices $\boldsymbol{\Sigma}_1$ and $\boldsymbol{\Sigma}_2$. If $\mathbf{J}_1$ is invertible at the true value $\boldsymbol{\theta}$ and $\|\ddot{\boldsymbol{\Phi}}_n(\boldsymbol{\theta}_1)\|_\mathrm{F}$ is $O_p(1)$ at some $\boldsymbol{\theta}_1$ between the estimator $\hat{\boldsymbol{\theta}}$ and the true $\boldsymbol{\theta}$, where $\|\ddot{\boldsymbol{\Phi}}_n(\boldsymbol{\theta}_1)\|_\mathrm{F} = \|\partial\mathrm{vec}[\mathbf{J}_1(\boldsymbol{\theta}_1)]/\partial\boldsymbol{\theta}_1^T\|_\mathrm{F}$, then $\sqrt{n}(\hat{\boldsymbol{\theta}} - \boldsymbol{\theta})$ converges in distribution to $N(0, \mathbf{K}\mathbf{J}_1^{-1}\boldsymbol{\Phi}_{\boldsymbol{\theta}}\mathbf{J}_1^{-1}\mathbf{K}^T)$, where $\boldsymbol{\Phi}_{\boldsymbol{\theta}}$ is a matrix of second-order limiting estimating functions given in the Supplement Section J.*

For the envelope model (4.3) with a non-normal error $\boldsymbol{\varepsilon}$, a good way to achieve parameter estimation is to use the same objective functions (4.4) and (4.5) to estimate the row and column envelopes, and then estimate the remaining parameters as described in Section 4.3. The next proposition shows that without normality on the error, the envelope estimators $g(\hat{\boldsymbol{\zeta}})$ obtained from this procedure still retain desired asymptotic properties.

**Proposition 6.** *Under (4.3) and the conditions in Proposition 5, $\sqrt{n}(g(\hat{\boldsymbol{\zeta}}) - g(\boldsymbol{\zeta}))$ converges in distribution to $N(0, \boldsymbol{\Lambda}^*)$, where $\boldsymbol{\Lambda}^* = \mathbf{K}\mathbf{G}_1(\mathbf{G}_1^T\mathbf{J}_1\mathbf{G}_1)^\dagger\mathbf{G}_1^T\boldsymbol{\Phi}_{\boldsymbol{\theta}}\mathbf{G}_1(\mathbf{G}_1^T\mathbf{J}_1\mathbf{G}_1)^\dagger\mathbf{G}_1^T\mathbf{K}^T$. Moreover, the envelope estimators are asymptotically more efficient than the estimators from (2.2) if $\mathrm{span}(\mathbf{J}_1^{1/2}\mathbf{G}_1)$ is a reducing subspace of $\mathbf{J}_1^{-1/2}\boldsymbol{\Phi}_{\boldsymbol{\theta}}\mathbf{J}_1^{-1/2}$.*

# 6 Sparse matrix-variate regression

While the matrix-variate regression can improve over conventional vector regression and the envelope model can gain further efficiency, they assume that all the rows and the columns of the matrix-valued response depend on the predictors. In application, especially for high dimensional data, it is possible that dependency might exist in a sparse manner. Therefore, it is meaningful to further study sparse matrix-variate regression and sparse envelope matrix-variate regression to accommodate such structure. Sparse regression models have been widely studied in multivariate settings (e.g. Turlach et al., 2005; Rothman et al., 2010; Chen and Huang 2012; Su et al., 2016). In particular, Su et al. (2016) studied sparse multivariate regression with an envelope structure. However, no investigation has been done in the matrix-variate setting. In the following work, we allow for the possibility that only certain variables in the response matrix are related to the predictors, leading to enveloping with simultaneous response variable selection, which can be useful when the matrix dimensions of $\mathbf{Y}$ are high.

Let $\mathcal{A}_1 \subseteq \{1, 2, \cdots, r\}$ be any index set with size $r_1 = |\mathcal{A}_1|$ ($r_1 < r$), then the rows of the response $\mathbf{Y}$ can be partitioned into two components $\mathbf{Y}_{\mathcal{A}_1,*}$ and $\mathbf{Y}_{\mathcal{A}_1^c,*}$. The part $\mathbf{Y}_{\mathcal{A}_1^c,*}$ is called inactive if

$$\mathbf{Y}_{\mathcal{A}_1^c,*}|\mathbf{X} \sim \mathbf{Y}_{\mathcal{A}_1^c,*}.$$

Similarly, let $\mathcal{A}_2 \subseteq \{1, 2, \cdots, m\}$ be any index set with size $m_1 = |\mathcal{A}_2|$ ($m_1 < m$). The columns of the response $\mathbf{Y}$ then can be partitioned into two components $\mathbf{Y}_{*,\mathcal{A}_2}$ and $\mathbf{Y}_{*,\mathcal{A}_2^c}$, and $\mathbf{Y}_{*,\mathcal{A}_2^c}$ is inactive if

$$\mathbf{Y}_{*,\mathcal{A}_2^c}|\mathbf{X} \sim \mathbf{Y}_{*,\mathcal{A}_2^c}.$$

Therefore, given the inactive row and column indices of $\mathbf{Y}$, the only active part of $\mathbf{Y}$ is $\mathbf{Y}_{\mathcal{A}_1,\mathcal{A}_2}$, which is affected by $\mathbf{X}$. Let $\boldsymbol{\beta}_{1,\mathcal{A}_1} \in \mathbb{R}^{r_1 \times p_1}$ be the matrix that contains the rows $\mathcal{A}_1$ of $\boldsymbol{\beta}_1$, and let $\boldsymbol{\beta}_{2,\mathcal{A}_2} \in \mathbb{R}^{m_1 \times p_2}$ be the matrix that contains the rows $\mathcal{A}_2$ of $\boldsymbol{\beta}_2$. Let $\boldsymbol{\beta}_{1,\mathcal{A}_1^c} \in \mathbb{R}^{(r-r_1) \times p_1}$ and $\boldsymbol{\beta}_{2,\mathcal{A}_2^c} \in \mathbb{R}^{(m-m_1) \times p_2}$ be the corresponding complements. Then under the sparse setting, both $\boldsymbol{\beta}_{1,\mathcal{A}_1^c}$ and $\boldsymbol{\beta}_{2,\mathcal{A}_2^c}$ are zero matrices.

Correspondingly, sparsity constraints can be added for parameter estimation based on (2.2). We employed the coordinate-independent penalty function (Chen et al., 2010; Su et al., 2016) that imposes a group lasso type of penalty (Yuan and Lin, 2006) with adaptive weights (Zou, 2006) on the likelihood function to introduce row-wise sparsity for $\boldsymbol{\beta}_1$ and $\boldsymbol{\beta}_2$. The parameters are then estimated by minimizing the objective function

$$\begin{aligned} l(\boldsymbol{\mu}, \boldsymbol{\beta}_1, \boldsymbol{\beta}_2, \boldsymbol{\Sigma}_1, \boldsymbol{\Sigma}_2) &= c + \frac{nr}{2} \log|\boldsymbol{\Sigma}_2| + \frac{nm}{2} \log|\boldsymbol{\Sigma}_1| \\ &+ \frac{1}{2} \sum_{i=1}^{n} \operatorname{tr}\{\boldsymbol{\Sigma}_2^{-1}(\mathbf{Y}_i - \boldsymbol{\mu} - \boldsymbol{\beta}_1 \mathbf{X}_i \boldsymbol{\beta}_2^T)^T \boldsymbol{\Sigma}_1^{-1}(\mathbf{Y}_i - \boldsymbol{\mu} - \boldsymbol{\beta}_1 \mathbf{X}_i \boldsymbol{\beta}_2^T)\} \\ &+ \lambda_1 \sum_{i=1}^{r} w_{1,i} \|b_{1,i}\|_2 + \lambda_2 \sum_{j=1}^{m} w_{2,j} \|b_{2,j}\|_2, \end{aligned}$$

where $c$ is a constant, $b_{1,i}$ denotes the $i$th row of $\boldsymbol{\beta}_1$, $b_{2,j}$ denotes the $j$th row of $\boldsymbol{\beta}_2$, $\lambda_1$ and $\lambda_2$ are the tuning parameters, $w_{1,i}$ and $w_{2,i}$ are adaptive weights, and $\|\cdot\|_2$ denotes the $L_2$ norm.

Under the envelope setting, let $\mathbf{L}_{\mathcal{A}_1}$ and $\mathbf{R}_{\mathcal{A}_2}$ be matrices that contain the rows $\mathcal{A}_1$ of $\mathbf{L}$ and the rows $\mathcal{A}_2$ of $\mathbf{R}$, respectively, and let $\mathbf{L}_{\mathcal{A}_1^c}$ and $\mathbf{R}_{\mathcal{A}_2^c}$ be the corresponding complements. Then $\mathbf{L}_{\mathcal{A}_1^c} = 0$ implies that $\boldsymbol{\beta}_{1,\mathcal{A}_1^c} = 0$ and $\mathbf{R}_{\mathcal{A}_2^c} = 0$ implies that $\boldsymbol{\beta}_{2,\mathcal{A}_2^c} = 0$, assuming that $u_1 \leq r_1$ and $u_2 \leq m_1$. Correspondingly, to achieve sparsity on the coefficient matrices, we impose the group lasso penalty on $\mathbf{L}$ and $\mathbf{R}$, respectively, for the envelope model. Therefore, for estimation we modify the objective functions in (4.4) and (4.5) to

$$f_{\mathbf{L}|2}(\mathbf{G}) = \log|\mathbf{G}^T \mathbf{S}_{\text{res}|2} \mathbf{G}| + \log|\mathbf{G}^T \mathbf{S}_{\mathbf{Y}|2}^{-1} \mathbf{G}| + \lambda_1 \sum_{i=1}^{r} w_{1,i} \|G_i\|_2, \qquad (6.1)$$

$$f_{\mathbf{R}|1}(\mathbf{U}) = \log|\mathbf{U}^T \mathbf{S}_{\text{res}|1} \mathbf{U}| + \log|\mathbf{U}^T \mathbf{S}_{\mathbf{Y}|1}^{-1} \mathbf{U}| + \lambda_2 \sum_{j=1}^{m} w_{2,j} \|U_j\|_2, \qquad (6.2)$$

where $G_i$ is the $i$th row of $\mathbf{G}$ and $U_j$ is the $j$th row of $\mathbf{U}$. We attain the minimization of $f_{\mathbf{L}|2}(\mathbf{G})$ and $f_{\mathbf{R}|1}(\mathbf{U})$ by a non-Grassman and blockwise coordinate descent algorithm (Su et al. 2016; Cook et al. 2016) to update $\mathbf{G}$ and $\mathbf{U}$ iteratively until convergence. Once $\mathbf{L}$ and $\mathbf{R}$ are estimated, the rest of the parameters are estimated in the same way as in the non-sparse setting. Numerical results for evaluating the sparse envelope matrix regression are given in the Supplement Section M.

The predictor variable selection can be similarly achieved by imposing group lasso type of penalty on the columns of $\boldsymbol{\beta}_1$ and/or $\boldsymbol{\beta}_2$, or lasso type of penalty on the individual elements of $\boldsymbol{\beta}_1$ and/or $\boldsymbol{\beta}_2$. The algorithm will be similar but will require more iterations for simultaneous response and predictor selection. We leave this part and theoretical properties for future study.

In high dimension settings, the full subset selection of envelope dimensions with AIC, BIC or cross validation can become relatively expensive. We thus propose stepwise selection methods for envelope dimension selection in high dimension. The methods along with numerical results are presented in the Supplement Section N.

# 7 Simulation studies

In this section, we evaluate the performance of matrix-variate regression (2.2) and envelope matrix-variate regression (4.3) numerically, and compare the two models along with the vector regression model (2.1). We first generated data based on (4.3) with $r = m = p_1 = p_2 = 5$, $u_1 = u_2 = 2$, $\boldsymbol{\Omega}_1 = \sigma^2 \mathbf{I}_{u_1}$, $\boldsymbol{\Omega}_{10} = \sigma_0^2 \mathbf{I}_{r-u_1}$, $\boldsymbol{\Omega}_2 = \sigma^2 \mathbf{I}_{u_2}$, and $\boldsymbol{\Omega}_{20} = \sigma_0^2 \mathbf{I}_{m-u_2}$. Here $\sigma^2 = .5$ and $\sigma_0^2 = 2.5$. The semi-orthogonal matrices $\mathbf{L}$ and $\mathbf{R}$ were generated by orthogonalizing matrices of independent uniform (0,1) random variables. The elements of $\boldsymbol{\mu}$, $\boldsymbol{\eta}_1$, $\boldsymbol{\eta}_2$ and the predictors were selected as independent standard random normal variables. We then fitted three models: (2.1), (2.2), and (4.3) to the data and evaluated their estimation accuracy of the coefficient parameters, according to the criteria: $|| \hat{\boldsymbol{\beta}}_2 \otimes \hat{\boldsymbol{\beta}}_1 - \boldsymbol{\beta}_2 \otimes \boldsymbol{\beta}_1 ||_F$ for the first two matrix regression models, and $|| \hat{\boldsymbol{\nu}} - \boldsymbol{\nu} ||_F$ for the vector regression (2.1). A corresponding evaluation of the covariance estimators showed results similar to the coefficient estimators and is thus omitted. We used six different sample sizes, $n = 200, 300, 500, 800, 1000, 1500$. Within each sample size, 200 replicates were simulated. The average estimation errors were computed over the 200 random samples for each sample size, under each model. The envelope dimensions were set to the true dimensions $u_1 = u_2 = 2$. With this setup models (2.1), (2.2) and (4.3) have 975, 105 and 75 real parameters to be estimated.

The left panel of Figure 2 shows a comparison of the three models, and the right panel highlights the improvement of enveloping. Over all selected sample sizes, the two matrix regression models provided much smaller estimation errors than did the vector regression. In addition, by effectively removing $X$-invariants from estimation, the envelope model further improved estimation accuracy in comparison to the matrix regression model without enveloping.

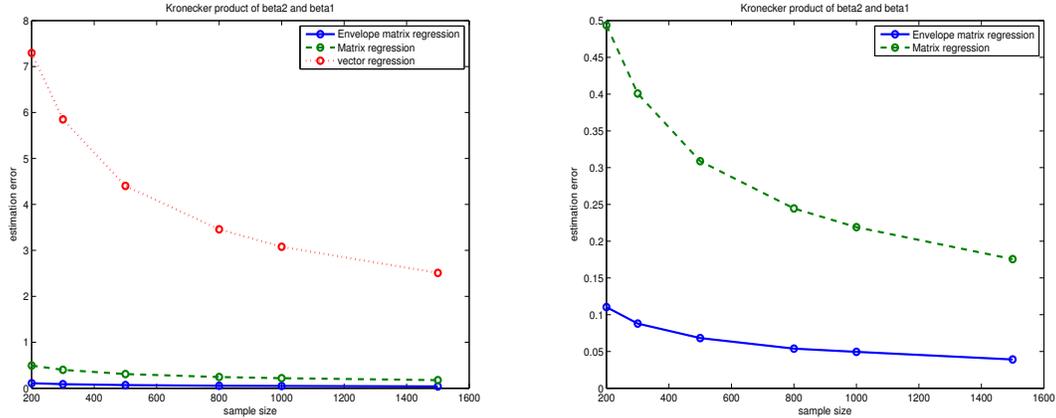

Figure 2: The average estimation errors of $\boldsymbol{\beta}_2 \otimes \boldsymbol{\beta}_1$ (or $\boldsymbol{\nu}$) from (2.1), (2.2), and (4.3).

Figure 3 shows the efficiency gains by assessing the asymptotic, actual, and bootstrap standard errors of the element in the first row and first column of $\hat{\boldsymbol{\beta}}_2 \otimes \hat{\boldsymbol{\beta}}_1$ from the three models. Similar results were obtained for other elements. The asymptotic standard errors were estimated according to the results presented in Propositions 1 and 2. The actual standard errors were computed using the sample standard errors of $\hat{\boldsymbol{\beta}}_2 \otimes \hat{\boldsymbol{\beta}}_1$ (or $\hat{\boldsymbol{\nu}}$) over 200 samples for each selected sample size. The bootstrap standard errors were obtained by bootstrapping one sample for each sample size. Again, the left panel in Figure 3 provides a comparison among the three models and the right panel highlights the comparison between the two matrix regression models with and without enveloping. Clearly, the asymptotic, actual and bootstrap standard errors are very close, indicating that the asymptotic covariance matrices proposed in Section 5 are accurate. In addition, over all selected sample sizes, the matrix regression without enveloping was asymptotically more efficient in parameter estimation than the vector regression, and imposing additional envelope structure further improved the efficiency gains. The importance of enveloping can be seen clearly in the right panel of Figure 3. Indeed, the ratios of the asymptotic standard errors between the estimators from (2.1) and (2.2) range from 23.7 to 54.9, while the ratios of the asymptotic standard errors between the estimators from (2.1) and (4.3) range from 309.0 and 523.9. Therefore, by utilizing matrix-variate information in model fitting, and at the same time, removing $X$-invariants from estimation, one can gain substantial efficiency in parameter estimation.

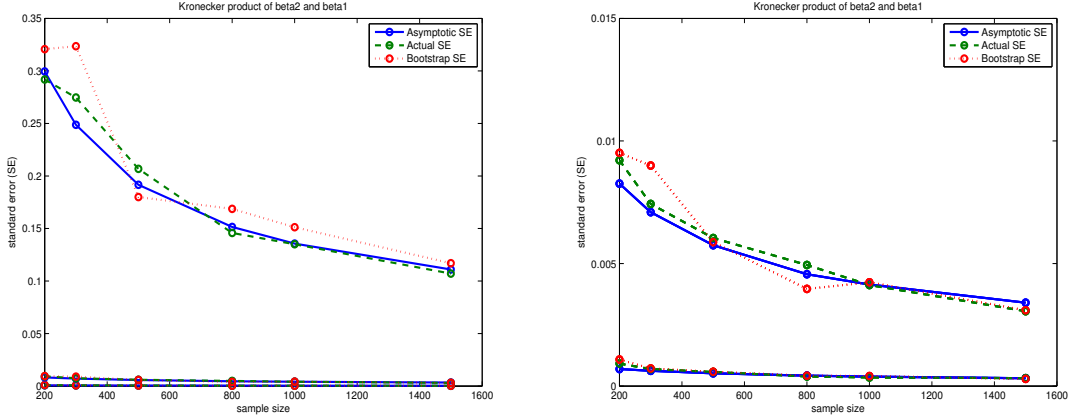

Figure 3: The standard errors of the element in the first row and first column of $\hat{\boldsymbol{\beta}}_2 \otimes \hat{\boldsymbol{\beta}}_1$ (or $\hat{\boldsymbol{\nu}}$) from the three models. In the left panel, the top three curves represent the corresponding standard errors from (2.1); the bottom three curves (that appear coincident with the horizontal axis) represent the corresponding results from the envelop model (4.3); the middle three curves represent the corresponding results from the matrix regression (2.2). In the right panel, the top three curves represent the corresponding standard errors from (2.2), and the bottom three curves represent the corresponding standard errors from (4.3).

We further assess the performance of the two matrix regression models (2.2) and (4.3) in terms of the estimation of individual row and column parameter matrices $\boldsymbol{\beta}_1$ and $\boldsymbol{\beta}_2$ in the Supplement Section K. The results regarding the estimation accuracy and estimation efficiency are shown in Figures 5 and 6 of the Supplement, which demonstrate similar patterns as the right panels of Figures 2 and 3. The improvement of the envelope estimation is substantial.

In Supplement Section O we give simulation results under misspecification of the envelope dimensions. The results indicate that envelopes still provide a mean squared error that is no greater than that for the matrix regression model and sometimes much smaller. In Supplement Sections L and M we present numerical results for matrix-variate regression, envelope matrix-variate regression, and sparse envelope matrix regression in high dimension, which demonstrate accuracy in both estimation and variable selection.

# 8 Applications

We applied the proposed matrix-variate regression models to the cross-over assay of insulin based on rabbit blood sugar concentration (Vϕlund, 1980) and the EEG data (Li et al. 2010). The design of the cross-over assay experiment along with the model fitting and results are presented in detail in the Supplement Section P, while we only include the results of the EEG data analysis in this section.

The EEG data was briefly introduced in the introduction section. It contains two groups of subjects: 77 subjects from the alcoholic group and 45 subjects from the control group. Each subject has measurements of electrical scalp activity, which form a $256 \times 64$ matrix. To explore the influence of the alcoholism on the brain activity, let $X = 1$ if a subject is alcoholic and $X = 0$ if a subject is nonalcoholic. We applied (B.2) to model the data as

$$\mathbf{Y} = \boldsymbol{\mu}_c + \boldsymbol{\beta} X + \boldsymbol{\varepsilon}, \tag{8.1}$$

where $\mathbf{Y} \in \mathbb{R}^{256 \times 64}$ is the matrix-valued brain measurements, $\boldsymbol{\mu}_c \in \mathbb{R}^{256 \times 64}$ is the mean brain activity of the nonalcoholic population (control group), $\boldsymbol{\beta} \in \mathbb{R}^{256 \times 64}$ is the difference between the mean brain activity of the alcoholic subjects, denoted as $\boldsymbol{\mu}_a$, and the mean brain activity of the nonalcoholic subjects, $\boldsymbol{\mu}_c$. Therefore, the coefficient matrix, $\boldsymbol{\beta} = \boldsymbol{\mu}_a - \boldsymbol{\mu}_c$, compares the alcoholic subjects and the nonalcoholic subjects based on their brain signals. The random error $\boldsymbol{\varepsilon}$ is assumed to follow $N(0, \boldsymbol{\Sigma}_1, \boldsymbol{\Sigma}_2)$. A saturated model such as (2.1) is not estimable for the EEG data as the dimension of vec($\mathbf{Y}$) is much higher than the sample size, and the the covariance matrix of vec($\mathbf{Y}$) given $X$ is not likely to be sparse. Matrix-variate regression provides one way to handle such data.

In (8.1), it is easy to see that the MLE of $\boldsymbol{\beta}$ without enveloping is $\bar{\mathbf{Y}}_a - \bar{\mathbf{Y}}_c$, the difference between the sample means of alcoholic brain activities and nonalcoholic brain activities. To achieve estimation efficiency, under the conditions (4.1) and (4.2), we considered fitting an envelope model on (8.1) as

$$\begin{aligned}\mathbf{Y} &= \boldsymbol{\mu}_c + \mathbf{L}\boldsymbol{\eta}\mathbf{R}^T X + \boldsymbol{\varepsilon} \\ \boldsymbol{\Sigma}_1 &= \mathbf{L}\boldsymbol{\Omega}_1\mathbf{L}^T + \mathbf{L}_0\boldsymbol{\Omega}_{10}\mathbf{L}_0^T, \quad \boldsymbol{\Sigma}_2 = \mathbf{R}\boldsymbol{\Omega}_2\mathbf{R}^T + \mathbf{R}_0\boldsymbol{\Omega}_{20}\mathbf{R}_0^T,\end{aligned} \tag{8.2}$$

where $\mathbf{L}$ and $\mathbf{R}$ are the bases of the $\boldsymbol{\Sigma}_1$-envelope of span($\boldsymbol{\beta}$) and the $\boldsymbol{\Sigma}_2$-envelope of span($\boldsymbol{\beta}^T$) respectively, $\boldsymbol{\eta}$ is the the coordinate matrix, and the rest of the parameters are similarly defined as in (4.3). The model estimation and inference of (8.2) are slightly different from those of (4.3). Details are provided in the Supplement Section Q. The BIC chose the envelope dimensions to be $u_1 = 7$ and $u_2 = 2$.

Similar to the results observed in the bioassay example, the envelope model can be more effective at identifying the differences between the alcoholic and nonalcoholic brain activities because it shows smaller standard errors for the estimators of $\boldsymbol{\beta}$. With $u_1 = 7$ and $u_2 = 2$, we calculated the relative ratios of the standard errors of the estimators from (8.1) to those from (8.2), and 90% of them fall in the range of 1.2 and 5.5. Moreover, since brain regions associated with alcoholism could be of interest to researchers, we further investigated the effect of alcoholism on different brain locations (channels) by averaging the two estimators of $\boldsymbol{\beta}$ over time (256 rows). We then obtained location-based estimators from (8.1) and (8.2), each of dimension 64. Bootstrap standard errors were computed for each element of the estimators and corresponding t-tests were applied. Because this is a multiple hypothesis testing problem, we employed the approach proposed by Benjamini and Yekutieli (2001) to control the false discovery rate (FDR; Benjamini and Hochberg, 1995). The FDR adjusted p-values (in $-\log_{10}$ scale) for the estimators with and without enveloping are reported in the left and right panels of Figure 4. The dashed lines in both panels represent the significance levels $-\log_{10}(0.05)$ and $-\log_{10}(0.1)$. Some plot characteristics are due to the adjustment process. For instance, about 25 p-values were adjusted to 1, which corresponds to $-\log_{10}(1) = 0$ in Figure 4. At the level $-\log_{10}(0.05)$, the envelope model suggests that 35 out of 64 brain regions are associated with alcoholism, while the model without enveloping was unable to identify relevant regions with only a few borderline significance.

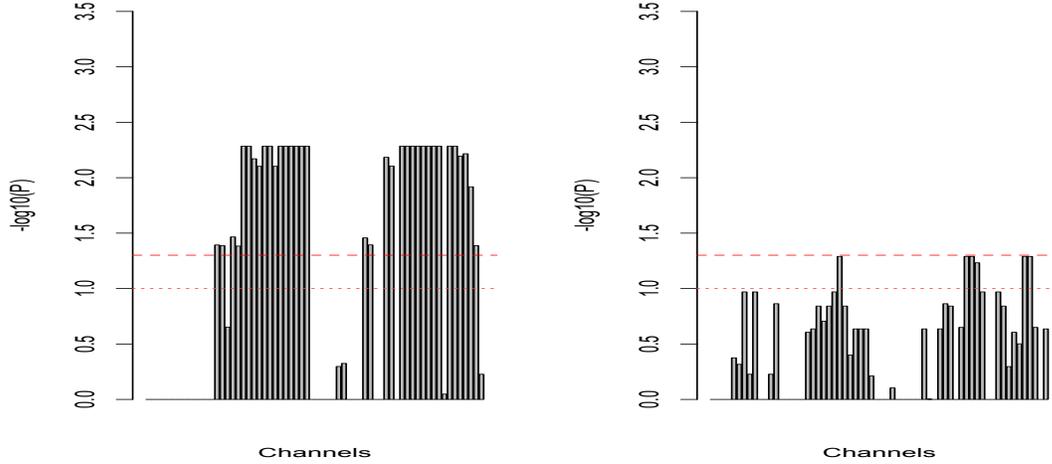

Figure 4: The FDR adjusted p-values (in $-\log_{10}$ scale) for the location-based estimators from the envelope model (8.2) (left panel) and the model without enveloping (8.1) (right panel).

Li and Zhang (2015) also analyzed these data, but computationally their method of estimation had to first reduce the number of time points from 256 to 64 by averaging adjacent groups of four. No such reduction was necessary with our estimation method. This is an important distinction. By averaging adjacent groups of four time points, a tuning parameter was essentially introduced into their estimation method. Different result would surely be obtained by varying the averaging group size. For instance, using BIC their method estimated envelope dimensions of $u_1 = 1$ and $u_2 = 1$, while using the full data we estimated $u_1 = 7$ and $u_2 = 2$. Averaging adjacent groups could lose dimensions.

# SUPPLEMENTARY MATERIALS

The supplementary materials contain proofs, technical details, and additional results.